\documentclass[sn-nature]{sn-jnl}

\usepackage{graphicx}%
\usepackage{multirow}%
\usepackage{amsmath,amssymb,amsfonts}%
\usepackage{amsthm}%
\usepackage{mathrsfs}%
\usepackage[title]{appendix}%
\usepackage{xcolor}%
\usepackage{textcomp}%
\usepackage{manyfoot}%
\usepackage{booktabs}%
\usepackage{algorithm}%
\usepackage{algorithmicx}%
\usepackage{algpseudocode}%
\usepackage{listings}%
\usepackage{braket}
\usepackage{etoolbox}

\makeatletter
\newcommand{\tel}[1]{\gdef\@tel{#1}}
\newcommand{\fax}[1]{\gdef\@fax{#1}}
\patchcmd{\@author}{\@email}{\@tel\par\@fax\par\@email}{}{}
\makeatletter

\theoremstyle{thmstyleone}%
\theoremstyle{thmstyletwo}%

\theoremstyle{thmstylethree}%

\begin{document}

\title[Article Title]{Comparative Analysis of Nanomechanical Resonators: Sensitivity, Response Time, and Practical Considerations in Photothermal Sensing}

\author[1]{\fnm{Kostas} \sur{Kanellopulos}}
\author[1]{\fnm{Friedrich} \sur{Ladinig}}
\author[1]{\fnm{Stefan} \sur{Emminger}}
\author[1]{\fnm{Paolo} \sur{Martini}}
\author[1]{\fnm{Robert} \sur{G. West}}
\author*[1]{\fnm{Silvan} \sur{Schmid}}\email{silvan.schmid@tuwien.ac.at}

\affil[1]{\orgdiv{Institute of Sensor and Actuator Systems}, \orgname{TU Wien}, \orgaddress{\street{Gusshausstrasse 27-29}, \city{Vienna}, \postcode{1040}, \country{Austria}}}


\abstract{Nanomechanical photothermal sensing has significantly advanced single-molecule/particle microscopy and spectroscopy, and infrared detection through the use of nanomechanical resonators that detect shifts in resonant frequency due to photothermal heating. However, the relationship between resonator design, photothermal sensitivity, and response time remains unclear. This paper compares three resonator types — strings, drumheads, and trampolines — to explore this relationship. Through theoretical modeling, experimental validation, and finite element method simulations, we find that strings offer the highest sensitivity (with a noise equivalent power of 280~fW/Hz$^{1/2}$ for strings made of silicon nitride), while drumheads exhibit the fastest thermal response. The study reveals that photothermal sensitivity correlates with the average temperature rise and not the peak temperature. Finally, the impact of photothermal back-action is discussed, which can be a major source of frequency instability. This work clarifies the performance differences and limits among resonator designs and guides the development of advanced nanomechanical photothermal sensors, benefiting a wide range of applications.}


\maketitle

\section{Introduction}\label{sec1}
In nanomechanical photothermal sensing, the resonator detects heat generated from various processes, including electromagnetic radiation absorption \cite{Zhang2013, Duraffourg2018, Blaikie2019, Vicarelli2022, Piller2023, Li2023, Zhang2024} and non-radiative energy transfer from minute samples \cite{Yamada2013, Biswas2014, Bose2014, Andersen2016, Kurek2017, Karl2018, Karl2020, Luhmann2023}, single molecules \cite{Chien2018}, single nanoparticles \cite{Larsen2013, Schmid2014, Ramos2018, Rangacharya2020, Chien2020a, Chien2021, Kanellopulos2023},  two-dimensional (2D) materials \cite{Kirchhof2022, Kirchhof2023}, and thin films \cite{Samaeifar2019, Cecacci2019}. 
Within this system, the nanomechanical resonator functions as the sensing element for the detection of energy exchange with the environment via resonance frequency shifts. As the resonator absorbs heat, its temperature rises, which decreases the tensile stress, leading to a corresponding frequency detuning. In other words, the resonator operates as a precise mechanical thermometer, offering significant advantages in terms of optimal sensor design and material choice \cite{West2023}. Notably, an optomechanical photothermal resonator fundamentally faces limitations from thermomechanical and photothermal back-action noise, in contrast to the electronic noise that plagues thermoelectric sensors and detectors \cite{rogalski2019infrared}.

\begin{figure*}
    \begin{center}
    \centering
    \includegraphics[width = 0.9\textwidth]{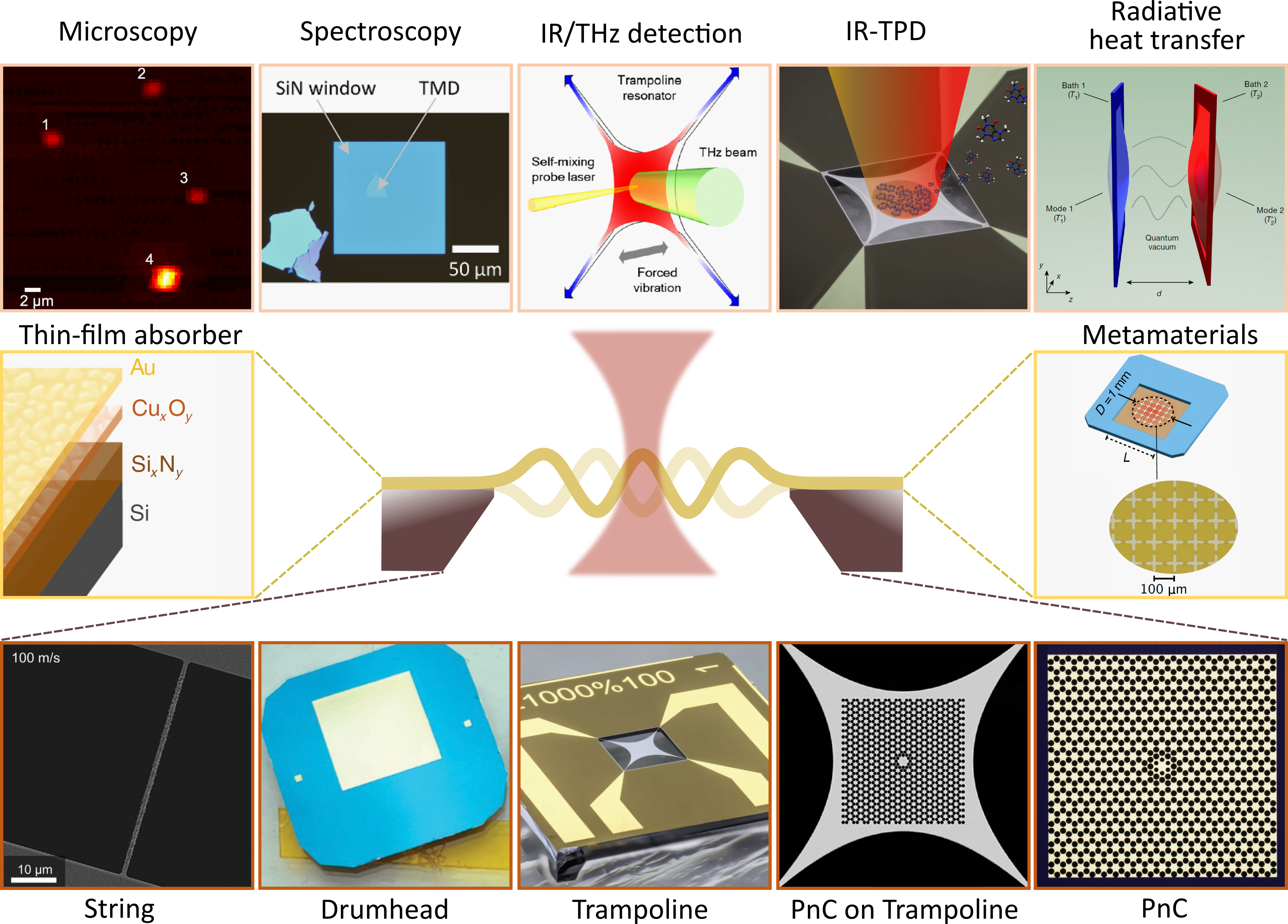}
    \caption{\textbf{Nanomechanical photothermal sensing.} To date, the photothermal effect in nanomechanical resonators has been explored in different fields of application (top row): molecular microscopy (reproduced and cropped from ref.~\citenum{Chien2018}--Copyright author(s) 2018, licensed under CC BY-NC-ND 4.0) and spectroscopy (reproduced and cropped from ref.~\citenum{Kirchhof2023}--Copyright 2023 under a CC BY 4.0 license), IR/THz detection (reproduced and cropped from ref.~\citenum{Vicarelli2022}--Copyright 2022 under a CC BY 4.0 license), IR-temperature programmed desorption (IR-TPD, reproduced and cropped from ref.~\citenum{Luhmann2023}--Copyright 2023 under a CC BY 4.0 license), and radiative heat transfer mechanisms (adapted with permission from ref.~\citenum{Fong2019}--Copyright 2019 by Springer Nature), among others. 
    Within this wealth of studies, different resonator designs have been used (bottom row): strings (reproduced and cropped from ref.~\citenum{Andersen2016}--Copyright 2016 under a CC BY-NC-ND 4.0 license), drumheads (adapted with permission from ref.~\citenum{Zhang2023}--Copyright 2023 by AIP Publishing), trampolines (reproduced and cropped from ref.~\citenum{Piller2023}--Copyright 2023 under a CC BY 4.0 license), phononic crystal (PnC) geometries on trampolines (reproduced and cropped from ref.~\citenum{Sadeghi2020pnc}--Copyright 2020 under a CC BY 4.0 license), and PnC alone (adapted with permission from ref.\citenum{Tsaturyan2017}--Copyright 2017 by Springer Nature).
    Depending on the application, addition of a further layer on top of the sensing area is also possible. Two examples are (central row): thin-film absorber (reproduced and cropped from ref.~\citenum{Luhmann2020}--Copyright 2020 under a CC BY 4.0 license), and metamaterials (reproduced and rearranged from ref.~\citenum{Zhang2024}--Copyright 2024 under a CC BY 4.0 license).}
    \label{fig:overview}
    \end{center}
\end{figure*}

Over the last decade, nanomechanical photothermal sensing has emerged as a powerful detection approach due to its high temperature sensitivity and versatility, as testified by the rapidly expanding areas of application, as outlined in Fig.~\ref{fig:overview}. To date, this technique has shown outstanding performances in the field of molecular microscopy and spectroscopy, operating in a wide range of the electromagnetic spectrum, from the visible \cite{Larsen2013, Chien2018, Rangacharya2020, Chien2020a, Chien2021}, near-infrared (IR) \cite{Kirchhof2022, Kirchhof2023, Kanellopulos2023}, to mid-IR \cite{Yamada2013, Biswas2014, Andersen2016, Kurek2017, Cecacci2019, Samaeifar2019, Luhmann2023}.  
In addition, the field is making important steps toward bridging the terahertz (THz) gap with resonant micro- and nanomechanical thermal detectors, offering a unique approach for room-temperature operation \cite{Zhang2013, Duraffourg2018, Blaikie2019, Piller2020, Zhang2020, Snell2022, Vicarelli2022, Piller2023, Zhang2023, Zhang2024}. Light-sound interaction in nanoresonators has been also successfully employed for enthalpy measurements \cite{Shakeel2018}, and detection of near-field heat radiation transfer \cite{Giroux2021, Giroux2023}, as well as phonon heat transfer through vacuum fluctuations \cite{Fong2019}.

So far, a variety of mechanical photothermal sensors has been employed for all this wealth of results, driven by different experimental requisites. 
For instance, silicon nitride string resonators have been extensively explored for photothermal sensing, in particular for microscopy and spectroscopy applications \cite{Larsen2011, Larsen2013, Yamada2013, Schmid2014, Bose2014, Biswas2014, Andersen2016, Karl2018, Karl2020, Rangacharya2020}. Their small cross-sectional area, together with the mechanical and thermal properties of SiN, make them extremely sensitive to temperature changes \cite{Larsen2011}.
With their high surface area, drumheads have been another platform of choice for nanomechanical photothermal spectroscopy of single nano-absorber \cite{Chien2018, Chien2021, Kanellopulos2023}, 2D materials \cite{Kirchhof2022, Kirchhof2023} and thin films \cite{Samaeifar2019, Cecacci2019}, as well as for detection of IR/THz radiation \cite{Blaikie2019, Piller2023, Giroux2021, Giroux2023, Zhang2024}. As it will be shown here, this advantage comes at the expense of a reduced thermal sensitivity.
Recently, trampoline resonators have also been employed in a variety of works in photothermal sensing \cite{Chien2020b, Pluchar2020, Vicarelli2022, Piller2023, Land2024}, due to their improved thermal insulation compared to drumheads, leading to competitive sensitivities with respect to the strings.
However, no comprehensive modeling and comparison of different resonator designs has been performed with respect to their sensitivity, response time, and practicality.

\begin{figure*}
    \begin{center}
    \centering
    \includegraphics[width = 0.8\textwidth]{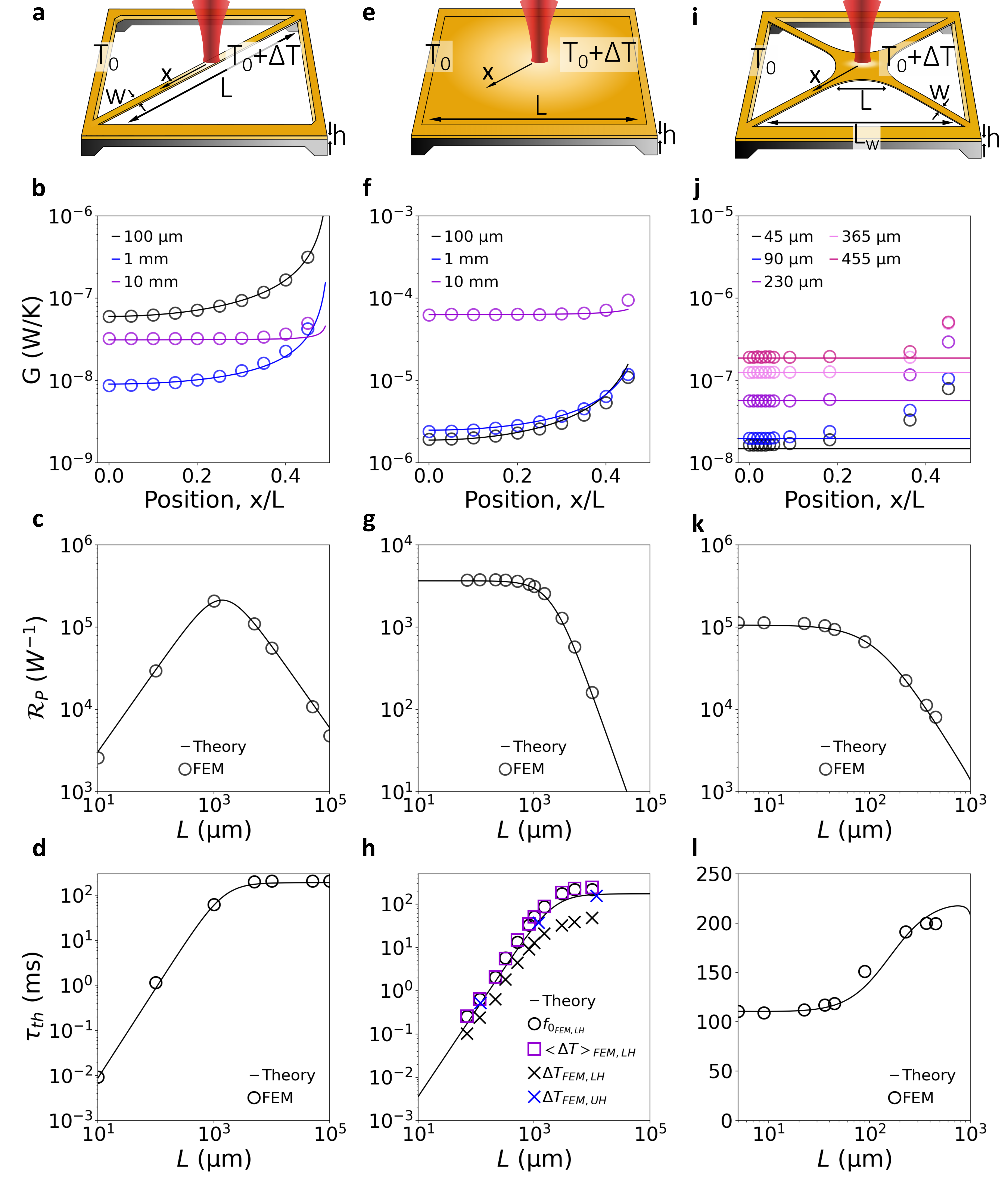}
    \caption{\textbf{FEM validation of MTF.} 
    \textbf{a.} Schematics of a string resonator illuminated by a light source (red) at the center. At thermal equilibrium, a temperature difference $\Delta T$ from the frame temperature $T_0$, will arise upon photothermal heating.  
    \textbf{b.} String's thermal conductance $G$ as a function of the point-like heat source relative position $x/L$, for three different lengths ($0.1$, $1$, and $10\ \mathrm{mm}$). Circles: FEM results of $G$ in the mean temperature framework (MTF). Solid curve: MTF theoretical calculation \eqref{eq_G_mtf}.
    \textbf{c.} Comparsion between FEM results (circles) and model (solid curve) for the relative power responsivity \eqref{eq_RP} as a function of the string length.
    \textbf{d.} Comparison between FEM results (circles) and model (solid curve) for the thermal time constant \eqref{eq_tauth} as a function of the string length.
    Model and FEM used parameters: $\rho = 3000\ \mathrm{kg/m^3}$, $c_p=700\ \mathrm{J/(kg\ K)}$, $\kappa = 3\ \mathrm{W/(m\cdot K)}$, $E = 250\ \mathrm{GPa}$, $\sigma_0=200\ \mathrm{MPa}$, $\nu=0.23$, $\alpha_\mathrm{th} = 2.2\ \mathrm{ppm/K}$, $\epsilon_\mathrm{rad} = 0.05$, $\alpha=0.5\ \%$, $w = 5\ \mathrm{\mu m}$, $h=50\ \mathrm{nm}$.
    \textbf{e.} Schematics of a drumhead resonator. 
    \textbf{f.} Circular drumhead's MTF thermal conductance for $0.1$, $1$, and $10\ \mathrm{mm}$ side length.
    \textbf{g.} Relative power responsivity comparison for drumheads.
    \textbf{h.} Thermal time constant comparison for drumheads. Solid curve: theory \eqref{eq_tauth}. Blue crosses: response time of the FEM peak temperature $\Delta T_{FEM}$ for uniform heating (UH). Black crosses: response time of $\Delta T_{FEM}$ for local heating (LH). Purple squares: response time of the surface mean temperature $\braket{\Delta T_{FEM}}$ for LH. Black circles: response time of the resonance frequency $f_{0_{FEM}}$ for LH.    Model and FEM used parameters: $\rho = 3000\ \mathrm{kg/m^3}$, $c_p=700\ \mathrm{J/(kg\ K)}$, $\kappa = 2.7\ \mathrm{W/(m\cdot K)}$, $E = 250$ GPa, $\sigma_0=50\ \mathrm{MPa}$, $\nu=0.23$, $\alpha_\mathrm{th} = 1.23\ \mathrm{ppm/K}$, $\epsilon_\mathrm{rad} = 0.05$, $\alpha=0.5\ \%$, $h=50\ \mathrm{nm}$.
    \textbf{i.} Schematics of a trampoline resonator.
    \textbf{j.} Trampolines' MTF thermal conductance for a frame window side length of $1.1$ mm and five different central pad side lengths. The model (solid curve) accounts here for a heat source impinging only in the central pad.
    \textbf{k.} Relative power responsivity comparison for trampolines.
    \textbf{l.} Thermal time constant comparison for trampolines. Model and FEM used parameters: $\rho = 3000\ \mathrm{kg/m^3}$, $c_p=700\ \mathrm{J/(kg\ K)}$, $\kappa = 2.7\ \mathrm{W/(m\ K)}$, $E = 250\ \mathrm{GPa}$, $\sigma_0=200\ \mathrm{MPa}$, $\nu=0.23$, $\alpha_\mathrm{th} = 2.2\ \mathrm{ppm/K}$, $\epsilon_\mathrm{rad} = 0.05$, $\alpha=0.5\ \%$, $w=5\ \mathrm{\mu m}$, $h=50\ \mathrm{nm}$ \cite{Piller2020, Sadeghi2020pnc}. For all the FEM simulations, a Gaussian beam of waist $w_0=1\ \mathrm{\mu m}$ has been used.
    }
    \label{fig:fem_model}
    \end{center}
\end{figure*}

This study establishes a comprehensive theoretical framework aimed at assessing the photothermal sensing performance of nanomechanical resonators, with a focus on noise equivalent power and response time. The analytical models illuminate, in particular, the interplay between sensor responsivity and frequency stability. Models for the individual noise components of the frequency stability are derived, including additive phase noise, temperature fluctuation noise, and photothermal back-action noise. The models herein are rigorously validated through comparison with experimental data and finite element method (FEM) simulations across varied nanomechanical silicon nitride resonator designs, namely strings, square drumheads, and trampolines, as schematically depicted in Fig.~\ref{fig:fem_model}a,e,I.

\section{Theory}
The figure of merit of a photothermal sensor is the noise-equivalent-power (NEP) with units [$\mathrm{W/\sqrt{Hz}}$], which for nanomechanical resonators is defined as \cite{Schmid2023}
\begin{equation}\label{eq_NEP}
    \text{NEP} = \frac{\sqrt{S_y(\omega)}}{\mathcal{R}_{\mathrm{P}}(\omega)},
\end{equation}
with  the one-sided spectral density fractional frequency fluctuations $S_y(\omega)$, with units [$\mathrm{1/Hz}$], and the relative power responsivity  $\mathcal{R}_{\mathrm{P}}(\omega)$, with units [$\mathrm{1/W}$].

The relative responsivity is defined as the fractional detuning of the resonator eigenfrequency $\omega_0$ per absorbed power $P$ and is given by \cite{Schmid2023}
\begin{equation} \label{eq_RP}
    \mathcal{R}_{\mathrm{P}}(\omega) = \frac{\partial \omega_0}{\partial \mathrm{P}}\frac{1}{\omega_0}\left\lvert H_{\mathrm{th}}(\omega)\right\rvert =\ \frac{\mathcal{R}_{\mathrm{T}}}{G}\left\lvert H_{\mathrm{th}}(\omega) \right\rvert,
\end{equation}
where $\mathcal{R}_{\mathrm{T}}$ denotes the relative temperature responsivity with units [$\mathrm{1/K}$], $G$ the thermal conductance [$\mathrm{W/K}$], and $H_{\mathrm{th}}(\omega) = (1 + i\omega\tau_\mathrm{th})^{-1}$ a low-pass filter transfer function accounting for the resonators' thermal response time \cite{Kruse2001a}
\begin{equation}  \label{eq_tauth}
    \tau_{\mathrm{th}} =\ \frac{C}{G},
\end{equation}
with $C$ denoting the resonators' heat capacity.
The absorbed power is the fraction of irradiated or otherwise introduced power $P_{0}$ that is converted into heat 
\begin{equation}
    P = \alpha P_0.
\end{equation}
with the absorber- and wavelength-dependent heat conversion factor $\alpha$ ($0 \le \alpha \le 1$).

\subsection{Temperature responsivity}
According to Eq.~\eqref{eq_RP}, nanomechanical photothermal sensors are, in essence, temperature sensors. The temperature responsivity is defined as
\begin{equation}\label{eq:RT}
    \mathcal{R_{\mathrm{T}}} = \frac{\partial \omega_0}{\partial T}\frac{1}{\omega_0}.
\end{equation}
The eigenfrequency of the resonators considered in this work is a function of the temperature-dependent tensile stress $\sigma(T)$
\begin{equation}\label{eq:omega0}
    \omega_0 \propto \sqrt{\sigma(T)},
\end{equation}
while the effect of bending stiffness is neglected.

\subsubsection{Strings}
In a string resonator with an intrinsic uniaxial tensile stress $\sigma_0$ and Young's modulus $E$, a mean temperature increase $\braket{\Delta T}$ induces a thermal strain along the resonator's length $L$, resulting in a stress \cite{Schmid2023}
\begin{equation} \label{eq_Stressstrg}
    \sigma(T) =\ \sigma_0 -\ E\ \alpha_{\mathrm{th}}\ \braket{\Delta T},
\end{equation}
with $\alpha_{\mathrm{th}}$ being the material's linear coefficient of thermal expansion. For small temperature changes, the temperature responsivity (\ref{eq:RT}) together with (\ref{eq:omega0}) and (\ref{eq_Stressstrg}) approximately is given by
\begin{equation} \label{eq_RTstrg}
    \mathcal{R}_\mathrm{T} = -\frac{\alpha_\mathrm{th}}{2}\frac{E}{\sigma_0}.
\end{equation}
The factor $E/\sigma_0$ is called the photothermal enhancement factor and is a unique feature of resonators under tensile stress. For nanomechanical silicon nitride resonators, the photothermal enhancement factor can reach values between $10^2 - 10^8$.

\subsubsection{Drumheads}
For very thin ($h \ll L$) homogeneous isotropic drumheads, the assumption of thin shell holds, and the thermal stress is given by \cite{Ventsel2001}
\begin{equation} \label{eq_Stressmbrn}
    \sigma(T) =\ \sigma_0\ -\ \frac{\alpha_{\mathrm{th}}\ E}{1\ -\ \nu}\braket{\Delta T},
\end{equation}
with $\nu$ being the resonator's Poisson's ratio. Hence, the relative temperature responsivity is
\begin{equation} \label{eq_RTmbrn}
    \mathcal{R}_{\mathrm{T}} =\ -\ \frac{\alpha_{\mathrm{th}}}{2\ (1\ -\ \nu)} \frac{E}{\sigma_0},
\end{equation}
with the factor ($1-\nu$) accounting for the thermal expansion along the two in-plane directions (for a detailed discussion about the dependence of $\mathcal{R}_T$ on the heat localization in drumheads, see SI Section S2).

\subsubsection{Trampolines}
The trampolines exhibit a thermal response similar to strings. For a central pad of area $L^2$ and thickness $h$, anchored to the frame via four tethers of length $L_{\mathrm{t}}$ and rectangular cross-section $w\cdot h$, its spring constant for the fundamental resonance mode can be expressed as (see SI Section S1)
\begin{equation} \label{eq_kteff}
    k_{\mathrm{eff}}(T) =\frac{\pi^2}{2} \frac{w h}{L_{\mathrm{t}}} (1-\nu) \sigma_0 \left [ 1 - \frac{\alpha_{\mathrm{th}} E}{\sigma_0} \braket{\Delta T} \right],
\end{equation}
with the factor $(1-\nu)$ accounting here for the strain release along the direction perpendicular to the tether length.
From the resonance frequency $\omega_0(T) \propto\sqrt{k_{\mathrm{eff}}}$, it is easy to observe that the temperature responsivity is equal to 
\begin{equation} \label{eq_RTtrmp}
    \mathcal{R}_\mathrm{T} = -\frac{\alpha_\mathrm{th}}{2}\frac{E}{\sigma_0},
\end{equation}
underlining that the thermal expansion at the tethers is the main responsible for the temperature response (see SI Section S1).

\subsection{Thermal conductance}
Besides the temperature responsivity, the power responsivity (\eqref{eq_RP}) also depends on the thermal conductance. 
The following thermal analysis is carried out based on the mean temperature framework (MTF) in the steady state that we introduce here.
The model is derived first assuming a point-like heat source. The case of an evenly spread heat source is discussed in the end of each subsection. 

A resonator of thermal mass $C$ absorbs a power $P$ and dissipates it to the environment through its conductance $G$, resulting in a mean temperature rise $\braket{\Delta T}$
\begin{equation}
    \braket{\Delta T} = \frac{P}{G}.
\end{equation}
In the MTF, all the resonator thermal properties are defined with respect to $\braket{\Delta T}$, as this temperature dictates the photothermal response of a nanomechanical resonator under tensile stress, rather than the local temperature variations $\Delta T$ (for a detailed discussion, see Fig.~\ref {fig:fem_model}h and SI Section S2). 

For an isotropic resonator, $C$ is given in the MTF by
\begin{equation} \label{eq_Cres}
    C =\ c_\mathrm{p}\ \rho\ V,
\end{equation}
where $c_p$, $\rho$, and $V$ are the specific heat capacity at constant pressure, mass density, and volume of the resonator, respectively.

As the resonator operates in a vacuum environment, only thermal conduction $ G_{\mathrm{cond}}$ and radiation $G_{\mathrm{rad}}$ contribute to the heat transfer \cite{Bergman2017}. 
In the MTF, the thermal conductance $G$ is given by
\begin{equation} \label{eq_G_mtf}
    G = G_{\mathrm{rad}} + G_\mathrm{cond} = 4 A_{\mathrm{rad}} \epsilon_{\mathrm{rad}} \sigma_{\mathrm{SB}} T_0^3 + \frac{s_\mathrm{f}(\mathbf{r},L, w_0)}{\beta(\mathbf{r}, L, w_0)} \kappa,
\end{equation}
where $A_{\mathrm{rad}}$, $\epsilon_{\mathrm{rad}}$, $\kappa$, and $\sigma_{\mathrm{SB}}$ are the resonator's radiating surface, its emissivity, thermal conductivity, and the Stefan-Boltzmann constant, respectively. For the thermal conduction term $G_\mathrm{cond}$, a shape factor $s_\mathrm{f}$ is introduced to account for the design geometry via the resonator characteristic length $L$, the heat source position vector $\mathbf{r}$, and the heating radius $w_0$ \cite{Bergman2017}. In this way, the dependence of $G$ on the size of the probing heat source, as well as on its position with respect to the resonator, e.g. concentric or eccentric to it, are taken into account within this formalism.
The product $s_\mathrm{f}(\mathbf{r}, L, w_0)\cdot \kappa$ is the thermal conduction with respect to the localized temperature field $\Delta T$. The factor $\beta(\mathbf{r}, L, w_0)$ denotes the ratio between mean and peak temperature $\beta = \braket{\Delta T} / \Delta T$, ensuring the correct description of $G_\mathrm{cond}$ in the MTF.

\subsubsection{Strings}
A string (Fig.~\ref{fig:fem_model}a) of length $L$, width $w$, and thickness $h$ occupies a volume $V = h w L$, and, assuming $h\ll w$, radiates with an area $A_\mathrm{rad}\approx 2 w L$. The factor of 2 accounts for the front and back surface radiation. This allows the direct evaluation of the thermal capacitance $C$ and radiative conductance $G_\mathrm{rad}$. For $G_\mathrm{cond}$ instead, shape $s_\mathrm{f}$ and $\beta$ factors have to be calculated. In this regard, it has to be noticed that the heating source can be in any position $x$ along the string length, with the generated heat flowing along two paths of length $x$ and $L-x$ \cite{West2023, Kruse2001a}. For a point-like heat source ($w_0 \xrightarrow{} 0$), $s_\mathrm{f}$ and $\beta$ are given by 
\begin{align}
    &s_\mathrm{f}(x) =\ \frac{4\ w\ h}{L}\ \frac{1}{1\ -\ \left(\frac{2\ x}{L}\right)^2}, \, \, \text{and}  \label{eq_Sfstrg} \\
    &\beta(x) =\ \frac{1}{2}. \label{eq_beta_strg}
\end{align}
Eq.~\eqref{eq_beta_strg} is valid as long as the temperature profile is linear, for which $\braket{\Delta T}=\Delta T / 2$ (see SI Section S2). 

Fig.~\ref{fig:fem_model}b shows the overall thermal conductance $G$ \eqref{eq_G_mtf} as a function of the relative localized heating position for three different strings. The MTF model (solid curves) closely aligns with the FEM simulations (circles), where the conductance has been extract as $G_\mathrm{FEM}=P/\braket{\Delta T_\mathrm{FEM}}$. For strings measuring 0.1 mm and 1 mm in length (black and blue curves, respectively), $G$ strongly depends on the heat source position, increasing as the latter approaches the frame, due to the enhanced thermal conduction. This effect is less pronounced for the $10$ mm long string, where radiative heat transfer dominates. It is worth noting that the $1$ mm long string shows the best thermal insulation, followed by the $10$ mm long and $0.1$ mm resonators, consistent with the theoretical and experimentally determined power responsivity $\mathcal{R}_\mathrm{P}$ (see Fig.~\ref{fig:fem_model}$c\&$\ref{fig:String_plots}c). 

For the case that the heating point source is located in the string center ($x=0$), $G_{\mathrm{cond}}$ can be expressed as
\begin{equation} \label{eq_Gcondstrg}
    G_{\mathrm{cond}} = \frac{s_\mathrm{f}(x=0)}{\beta(x=0)}\ \kappa =\ \frac{8\ h\ w}{L}\ \kappa.
\end{equation}
The MTF predicts a factor of 2 higher than what is reported in ref.~\citenum{Schmid2023}, as $\braket{\Delta T}$ is considered instead of the peak temperature. 

Fig.~\ref{fig:fem_model}c compares the theoretical power responsivity (\ref{eq_RP}) to the FEM simulations, showing excellent agreement. For short lengths ($L<1\ \mathrm{mm}$), $\mathcal{R}_\mathrm{P}$ increases linearly with $L$, until it reaches a maximum ($L \approx 1$ mm). In this region, the string is fully coupled to the thermal bath via thermal conduction ($G=G_\mathrm{cond}$). As the distance between the impinging and anchoring points increases, so does the power responsivity. For $L>1\ \mathrm{mm}$, the string enters the radiation limited regime ($G=G_\mathrm{rad}$), resulting in a linear reduction of $\mathcal{R}_\mathrm{P}$, due to the increasingly larger emitting surface area $A_\mathrm{rad}\propto L$. This comparison proves the validity of the responsivity model.

The thermal conductance \eqref{eq_G_mtf} also depends on the spot size of the probing laser. FEM simulations for a uniformly heated string shows that thermal insulation is $1.5\times$ less than under localized heating conditions. Thus, localizing the heat source at the string center will lead to a $1.5\times$ higher power response (see SI Section S2). 

Fig.~\ref{fig:fem_model}d displays the comparison between FEM (circles) and theoretical response time \eqref{eq_tauth}. Short strings are dominated by conductive heat transfer, with $\tau_\mathrm{th}$ being a linear function of $L$. Conversely, long strings are dominated by radiative heat transfer and show a time constant independent of $L$, as both the thermal capacitance $C$ and the conductance $G=G_\mathrm{rad}$ grow linearly with $L$. As can be observed, the model accurately predicts the string's time constant.
Notably, there is a trade-off between power responsivity and thermal time constant: for $L\leq 1\ \mathrm{mm}$ higher responsivity corresponds to a slower thermal time response.

\subsubsection{Drumheads}
A square drumhead resonator of side length $L$  and thickness $h$ (Fig.~\ref{fig:fem_model}e) has a volume $V= h L^2$ and a radiating surface $A_{\mathrm{rad}} = 2 L^2$. Given $h\ll L$, no thermal gradient is present along the thickness, and the eccentric shell model applies for the correct description of $G_\mathrm{cond}$ \cite{Bergman2017}, with the heat being dissipated isotropically to the frame. For simplicity, the model focuses on a circular drumhead of effective diameter $D = 2 L/\sqrt{\pi}$. For this geometry, the shape $s_\mathrm{f}$ and $\beta$ factors are given by (for the derivation, see SI Section S2)
\begin{align}
    &s_\mathrm{f}(\mathbf{r}, D, w_0) =\ \frac{4\ \pi\ h}{2\ cosh^{-1}\left[\frac{D^2\ +\ (2\ w_0)^2\ -\ 4\ r^2}{2\ D\ (2\ w_0)}\right] + 1}, \label{eq_Sfmbrn}\\
    &\beta(\mathbf{r},D,w_0) =\ \frac{1-\frac{1}{2}\left(\frac{2\ w_0}{D}\right)^2}{1 - 2\ \mathrm{ln}\left(\frac{2\ w_0}{D}\right)} \left[1 - \left(\frac{r}{D}\right)^2\right]. \label{eq_beta_mbrn}
\end{align}
Fig.~\ref{fig:fem_model}f shows the overall conductance $G$ as a function of a localized heat source position, for three different circular membranes ($L=0.1$, $1$, and $10$ mm). The MTF model (solid curves) closely aligns with the FEM simulations for circular drumheads (circles). The two smaller drumheads ($L<1$ mm, black and blue curves), primarly coupled to the environment via conduction, exhibit similar values. Conversely, the larger drumhead in the radiative heat transfer regime has a constant and worse thermal conductance. 

For a focused heat source at the drumhead center ($r=0$, $w_0\rightarrow 0$), the conductance becomes
\begin{equation} \label{eq_Gcondmbrn}
    G_{\mathrm{cond}} = \frac{s_\mathrm{f}(r=0)}{\beta(r=0)}\ \kappa=\ 4\ \pi\ h\ \kappa,
\end{equation}
recovering the same result of ref.~\citenum{Kurek2017}. Even in the case of a localized heat source, thermal conduction in drumheads is independent of the side length $L$, contrary to what happens in strings \eqref{eq_Gcondstrg}. 

Fig.~\ref{fig:fem_model}g shows the comparison between the theoretical (black solid curve) and FEM power responsivity (black circle), showing excellent agreement. Small drumheads ($L<1\ \mathrm{mm}$) shows a responsivity independent of $L$, being $G=G_\mathrm{cond}$ exclusively a function of the material thermal conductivity $\kappa$ and the resonator's thickness $h$ \eqref{eq_Gcondmbrn}. Large drumheads ($L>1\ \mathrm{mm}$) enter the radiative regime, and the responsivity drops down due to the increased surface area. This comparison confirms the validity of the theoretical responsivity model for drumhead resonators.

Drumheads show a different dependence on heat localization compared to strings. FEM simulations for a concentric Gaussian beam of varying waist $w_0$ have shown that the power responsivity $\mathcal{R}_P(w_0=L/2)\approx\mathcal{R}_P(w_0\ll 1) /2$, i.e. for a uniform heating condition (see SI Section S2). As a simple rule here, a point-like heat source offers a $2\times$ improved photothermal responsivity compared to uniformly distributed heating.

Fig.~\ref{fig:fem_model}h compares the theoretical and FEM modeling of the thermal time constant. The study considers uniform (UH) and local heating (LH) conditions. The theoretical predictions (black curve) closely align with the scenario of uniform light illumination (blue crosses), with $\tau_\mathrm{th}$ derived from the temporal evolution of the resonator's maximum temperature $\Delta T$. Notably, the thermal equilibrium is reached faster in the case of local heating (black crosses). For the same scenario, $\tau_\mathrm{th}$ has been additionally estimated through a transient study of the resonance frequency (black circles), revealing a stronger agreement with the theory. Monitoring the mean temperature $\braket{\Delta T}$ (dark purple squares) further supports this result: the two sets of FEM perfectly overlap, indicating that the resonance frequency is governed by the resultant mean temperature distribution even in the presence of a local heating source.
Opposite to what has been seen for strings, the most responsive drumheads show the fastest time response.

\subsubsection{Trampolines}
A trampoline (Fig.~\ref{fig:fem_model}i) occupies a volume $V=h \left(L^2 + 4 w L_{\mathrm{t}}\right)$ and radiates through its central pad and tethers with an area $A_{{\mathrm{rad}}} = 2 \left(L^2 + 4 w L_{\mathrm{t}}\right)$. The 2D heat conduction problem simplifies here to a 1D scenario, as in strings. Indeed, heat generated in any position on the central pad conductively dissipates through the tethers. Since the heat flow is constricted by the tethers, the resonator can be modeled as a cross-string. According to (\ref{eq_Sfstrg}), the resulting shape and $\beta$ factors are given by
\begin{align}
    &s_\mathrm{f}(x) =\ 2\ \frac{4\ w\ h}{2\ L_\mathrm{t}} \frac{1}{1 - \left(\frac{x}{L_\mathrm{t}}\right)^2},\,\, \text{and} \label{eq_Sftrmp}\\
    &\beta(\mathbf{r},L,w_0) =\ 1. \label{eq_beta_trmp}
\end{align}
The factor of 2 in Eq.~\eqref{eq_Sftrmp} accounts for the two crossing strings, while Eq.~\eqref{eq_beta_trmp} is defined only with respect to the central pad, being the core sensing area. 

Fig.~\ref{fig:fem_model}j shows the FEM computed values for $G$ for five trampolines of different central areas $L^2$ (circles), together with the MTF predictions (solid curves). As the heat source moves from the center to the frame along a tether, the thermal conduction $G_\mathrm{cond}$ increases. Moreover, both $G_\mathrm{cond}$ and $G_\mathrm{rad}$ rise for increasing area--the former due to shorter tethers, the latter due to a larger surface. For a tightly focused beam at the center, the thermal conductance is  
\begin{equation} \label{eq_Gcondtrmp}
    G_\mathrm{cond} = 4\ \frac{h\ w\ \kappa}{L_{\mathrm{t}}},
\end{equation}
recovering the results of a cross-string resonator of different tether lengths.

Fig.~\ref{fig:fem_model}k displays the theoretical and FEM simulated power responsivity as a function of the central side length $L$. The model aligns closely with FEM simulations: resonators with small areas ($L^2<100^2\ \mathrm{\mu m}^2$) show an almost constant $\mathcal{R}_\mathrm{P}$; for larger trampolines ($L^2>100^2\ \mathrm{\mu m}^2$), it decreases linearly as the pad area grows. The trend is similar to the drumhead case (Fig.~\ref{fig:fem_model}g). The difference in orders of magnitude compared to the drumheads relates to the improved thermal insulation (see Fig.~\ref{fig:fem_model}f$\&$j). As the window size is kept fixed, the growth of the central area corresponds to a reduction in tethers' length. For $L<100\ \mathrm{\mu m}$, long tethers provide high thermal insulation, with $\mathcal{R}_\mathrm{P}$ converging to the cross-string case. As $L^2$ approaches $L_\mathrm{w}^2$, thermal radiation, as well as conduction increases due to the tethers shortening, with $\mathcal{R}_\mathrm{P}$ approaching the drumhead performances. This comparison shows the validity of the thermal model employed so far.

As only the central pad is here the sensing area, uniform heating would result in an almost identical mean temperature rise $\braket{\Delta T}$ for this geometry, leading to no reduction of the power responsivity $\mathcal{R}_P(w_0\rightarrow 0)\approx \mathcal{R}_P(w_0=L/2)$ (see SI Section S2).

Fig.~\ref{fig:fem_model}l shows the thermal time constant comparison between the model (solid curve) and the FEM simulations (circles), showing excellent agreement. For $L<50\ \mathrm{\mu m}$, the trampoline behaves as a string. For $50\ \mathrm{\mu m}<L<230\ \mathrm{\mu m}$, the resonator thermal capacitance grows faster than the conductance, increasing the overall response time. For $L>230\ \mathrm{\mu m}$, $\tau_\mathrm{th}$ reaches a plateau, to drop down for increasingly larger central pads. This is explained by the increase in conduction and radiation: the former, due to the shorter tether length; the latter, due to a bigger central area $L^2$. The interplay between thermal mass and conductance is the same one observed between the effective mass and the stiffness for the resonance frequency, as shown in Fig.~\ref{fig:Trmp_plots}a (see also SI Section SI).

As for drumheads, the most responsive trampolines exhibit the fastest time response.

\begin{table*}\label{tab:Rt_G}
\caption{Expressions for the relative temperature responsivity $\mathcal{R}_\mathrm{T}$ and thermal conductance $G$ for the three designs. The two quantities are used to calculate the relative power responsivity \eqref{eq_RP}, for localized (LH) and uniform (UH) heating. For the drumheads see also SI Section S2.}
\begin{tabular}{|r|c|c|c|}
\hline
 & $\mathcal{R}_\mathrm{T}$  [1/K] & LH: $G$ [W/K] & UH: $G$ [W/K]\\
\hline
\hline
String & $-\frac{\alpha_\mathrm{th}}{2}\frac{E}{\sigma_0}$ & $8 \frac{w h}{L} \kappa + 8 w L \epsilon_\mathrm{rad} \sigma_\mathrm{SB} T_0^3$ & $12 \frac{w h}{L} \kappa + 8 w L \epsilon_\mathrm{rad} \sigma_\mathrm{SB} T_0^3$ \\
\hline
Drumhead & $-\frac{\alpha_\mathrm{th}}{2 (1 - \nu)}\frac{E}{\sigma_0}$ & $4 \pi h \kappa + 4 L^2 \epsilon_\mathrm{rad} \sigma_\mathrm{SB} T_0^3$ & $8 \pi h \kappa + 8 L^2 \epsilon_\mathrm{rad} \sigma_\mathrm{SB} T_0^3$\\ 
\hline
Trampoline & $-\frac{\alpha_\mathrm{th}}{2}\frac{E}{\sigma_0}$ & $8 \frac{w h}{L_\mathrm{t}} \kappa + 4(8w L_\mathrm{t} + 2 L^2) \epsilon_\mathrm{rad} \sigma_\mathrm{SB} T_0^3$ & $8 \frac{w h}{L_\mathrm{t}} \kappa + 4(8w L_\mathrm{t} + 2 L^2) \epsilon_\mathrm{rad} \sigma_\mathrm{SB} T_0^3$ \\
\hline
\end{tabular}
\end{table*}
A summary of the expressions of $\mathcal{R}_\mathrm{T}$ and $G$ for the calculation of the power responsivity \eqref{eq_RP} is displayed in Table~\ref{tab:Rt_G}, for point-like source and even illumination.

\subsection{Frequency stability}
High photothermal sensitivity (\ref{eq_NEP}) requires also low fractional frequency noise, as it defines the smallest resonance frequency shift that can be resolved. In nanomechanical photothermal sensing, the most relevant noise sources are: i) additive phase noise $S_{y_{\theta}}(\omega) = S_{y_{\theta_\mathrm{thm}}}(\omega)  + S_{y_{\theta_\mathrm{det}}}(\omega)$, sum of thermomechanical and detection noise \cite{Besic2023}; ii) temperature fluctuation frequency noise $S_{y_\mathrm{th}}(\omega)$ \cite{Zhang2023}; and iii) photothermal back-action frequency noise $S_{y_\mathrm{\delta P}}(\omega)$
\begin{equation} \label{eq_Sy}
    S_y(\omega) = S_{y_{\theta}}(\omega)  + S_{y_\mathrm{th}}(\omega) + S_{y_\mathrm{\delta P}}(\omega).
\end{equation}

\subsubsection{Additive phase noise}
Additive phase noise originates from the conversion of thermomechanical $S_{z_\mathrm{thm}}(\omega)$ and detection $S_{z_\mathrm{det}}(\omega)$ amplitude noise into phase noise \cite{Besic2023}. In the assumption of detection of white noise, this contribution can be expressed with respect to the thermomechanical noise peak as \cite{Schmid2023}
\begin{equation}\label{eq_Kd}
     S_{z_\mathrm{det}}(\omega) = \mathcal{K}_\mathrm{d}^2 S_{z_\mathrm{thm}}(\omega_0) = \mathcal{K}_\mathrm{d}^2 \left[\frac{4 k_\mathrm{B} T Q}{m_\mathrm{eff}\omega_0^3}\right]
\end{equation}
with $\mathcal{K}_{\mathrm{d}} < 1$ for transduction systems able to resolve the thermomechanical noise.
Assuming that the resonator is made to oscillate at an amplitude $z_\mathrm{r}$ by means of a closed-loop frequency tracking scheme, the resulting fractional frequency noise power spectral density (PSD) is \cite{Schmid2023}
\begin{equation}\label{eq_Sy_thm_closedloop}
    S_{y_{\theta}}(\omega) = \frac{1}{2Q^2}\frac{S_{z_\mathrm{thm}}}{z^2_\mathrm{r}}\left[ \lvert H_{\theta_\mathrm{thm}}(\mathrm{i} \omega)\rvert^2 + \mathcal{K}_\mathrm{d}^2 \lvert H_{\theta_\mathrm{det}}(\mathrm{i} \omega)\rvert^2   \right].
\end{equation}
$H_{\theta_\mathrm{thm}}(\mathrm{i} \omega)$ and $H_{\theta_\mathrm{det}}(\mathrm{i} \omega)$ are the loop-specific transfer functions for the thermomechanical and detection phase noise. The transfer functions for an open loop, phase-locked loop, and self-sustaining oscillator are the same to a good approximation \cite{Besic2023}. As an example, for a self-sustaining oscillator (SSO) scheme, as used in this work, the transfer functions are \cite{Besic2023}
\begin{equation}\label{eq_SSO}
\begin{aligned}
    &H_{\theta_\mathrm{thm}}^\mathrm{SSO}(\mathrm{i}\omega) = H_\mathrm{L}(\mathrm{i}\omega), \\
    &H_{\theta_\mathrm{det}}^\mathrm{SSO}(\mathrm{i}\omega) = \frac{H_\mathrm{L}(\mathrm{i}\omega)}{H_\mathrm{mech}(\mathrm{i}\omega)}.
\end{aligned}
\end{equation}
$H_\mathrm{mech}(\mathrm{i}\omega)$ and $H_\mathrm{L}(\mathrm{i}\omega)$ are the low-pass filter transfer functions of the resonator and system filter, respectively 
\begin{equation}
\begin{aligned}
    &H_\mathrm{mech}(\mathrm{i}\omega) = \frac{1}{1+\mathrm{i}\omega\tau_\mathrm{mech}}, \\
    &H_\mathrm{L}(\mathrm{i}\omega) = \frac{1}{1+\mathrm{i}\omega\tau_\mathrm{L}}.
\end{aligned}
\end{equation}
with the resonator time constant $\tau_\mathrm{mech}=2Q/\omega_0$ and the filter time constant $\tau_\mathrm{L} = 1 /(2\pi f_\mathrm{L})$. 

Additive phase noise \eqref{eq_Sy_thm_closedloop} can be mitigated by actuating the resonator at the onset of nonlinearity $z_{\mathrm{r_c}}$, 
\begin{equation} \label{eq_zc}
    z_{{\mathrm{r_c}}} =\ \sqrt{\frac{8}{3\ \sqrt{3}}}\frac{1}{\sqrt{Q}}\sqrt{\frac{m_{{\mathrm{eff}}}\ \omega_0^2}{\alpha_{{\mathrm{Duff}}}}},
\end{equation}
with $\alpha_{{\mathrm{Duff}}}$ denoting the effective Duffing term \cite{Schmid2023}. For $z_{\mathrm{r}}>z_{\mathrm{r_c}}$, additional phase noise of nonlinear origin could enter the system, worsening the resonator frequency stability at the integration times of interest in this study \cite{Manzaneque2023}. 

\subsubsection{Temperature fluctuation frequency noise}
Thermal fluctuation fractional frequency noise can also be assumed white. For a lumped-element model, $S_{y,\mathrm{th}}(\omega)$ is given by \cite{Vig1999, Kruse2001a}
\begin{equation} \label{eq_Syth}
    S_{y_\mathrm{th}}(\omega) =\ \frac{4\ k_B\ T^2}{G_{\mathrm{eff}}} \mathcal{R}_{\mathrm{T}}^2 \left\lvert \frac{1}{1 + i \omega \tau_\mathrm{th_{eff}}} \right\rvert^2.
\end{equation}

Here, $G_{\mathrm{eff}}$ and $\tau_\mathrm{th_{eff}}$ represent an effective thermal conductance and time constant, accounting for the fact that temperature fluctuations occur randomly at any point of the resonator. $G_\mathrm{eff}$ is derived from the integration of the conductance $G$ over all possible positions of a point-like heat noise source. Since radiation is heat source position-independent in MTF, only the integration of $G_\mathrm{cond}$ is required. From $G_\mathrm{eff}$, $\tau_\mathrm{th_{eff}} = C/ G_\mathrm{eff}$ can be evaluated.

In a string resonator, thermal noise can enter the system at any point along its length $L$. 
Integrating Eq.~\eqref{eq_Sfstrg} along $L$ gives the effective conductance 
\begin{equation} \label{eq_Geffstrg}
\begin{aligned}
    &G_{\mathrm{eff}} = \left( \frac{1}{\kappa}\frac{1}{L}\int_0^L{\frac{1}{s_\mathrm{f}(x)}}\text{d}x\right)^{-1} + 8 w L \epsilon_\mathrm{rad} \sigma_\mathrm{SB} T_0^3 = \\
    &=  \frac{6 \kappa w h}{L} + 8 w L \epsilon_\mathrm{rad} \sigma_\mathrm{SB} T_0^3.
\end{aligned}
\end{equation}
Eq.~\eqref{eq_Geffstrg} results in a higher conductance than \eqref{eq_Gcondstrg}, as the averaging includes noise sources closer to the clamping points, where $G_\mathrm{cond}$ increases exponentially (see Fig.~\ref{fig:fem_model}b).

For a circular drumhead, the integration over all the possible noise source positions gives
\begin{equation} \label{eq_Geffmbrn}
\begin{aligned}
    &G_\mathrm{eff} = \left(\frac{1}{\pi\ (\frac{D}{2})^2\ \kappa}\int_0^{2\pi}\int_{0}^{(D - d)/2}{\frac{1}{s_\mathrm{f}(r,\theta, D, d)}r}\text{d}r\text{d}\theta \right)^{-1}\\
    &+ 8 L^2 \epsilon_{\mathrm{rad}} \sigma_{\mathrm{SB}} T_0^3=\\
    &\simeq 4 \pi h \kappa + 8 L^2 \epsilon_{\mathrm{rad}} \sigma_{\mathrm{SB}} T_0^3.
\end{aligned}
\end{equation}
As with strings, the most important noise contribution is given by the central region, resulting in $G_\mathrm{eff}\simeq G(\mathbf{r}=0)$.

For a trampoline, random energy exchange can occur in both its central pad and along its four tethers, resulting in
\begin{equation} \label{eq_Gefftrmp}
\begin{aligned}
    &G_\mathrm{eff} = \left(\frac{1}{\kappa}\frac{1}{\sqrt{2}\ L_\mathrm{w}}\int_0^{\sqrt{2}L_w}{\frac{1}{s_\mathrm{f}(x)}}\text{d}x \right)^{-1} \\
    &+ 4(8w L_\mathrm{t} + 2 L^2) \epsilon_\mathrm{rad} \sigma_\mathrm{SB} T_0^3 = \\
    &= \frac{6\ \sqrt{2}\ \kappa\ w\ h}{L_\mathrm{w}} + 4(8w L_\mathrm{t} + 2 L^2) \epsilon_\mathrm{rad} \sigma_\mathrm{SB} T_0^3.
\end{aligned}
\end{equation}
with $L_\mathrm{w}$ denoting the window side length. While trampolines dissipate $\sqrt{2}\times$ more than strings via conduction, the central pad will make this geometry extremely sensitive to temperature fluctuations.

\subsubsection{Photothermal back-action frequency noise}

Photothermal back-action frequency noise $S_{y,\mathrm{\delta P}}(\omega, \lambda)$ originates from the intensity fluctuations of the light source employed for photothermal sensing, as well as any other light source used for transduction, such as interferometric lasers. For a continuous wave (CW) source with an intensity fluctuation PSD $S_I(\omega, \lambda)$ [$\mathrm{W^2/Hz}$] (also called relative intensity noise, RIN - see Fig.~\ref{fig:Sxx_mbrn_P_comp}c), the resonator fractional frequency fluctuations are given by
\begin{equation} \label{eq_Sypb}
    S_{y, \mathrm{\delta P}}(\omega, \lambda) =\ \alpha^2(\lambda)\ \mathcal{R}_\mathrm{P}^2(\omega)\ S_{I}(\omega, \lambda),
\end{equation}
where $S_I(\omega, \lambda)$ typically has the form
\begin{equation}\label{eq_Si}
    S_I(\omega, \lambda)=h_0+h_{-1}f^{-1}+h_{-2}f^{-2}
\end{equation}
for a generic laser source \cite{Maddaloni2013}. Here, $h_0$ denotes the laser shot-noise limit $S_{I,\mathrm{sn}}(\lambda)=2 h c \braket{P_0} / \lambda$, where $\braket{P_0}$ is the average input power; the terms $h_{-1}$ and $h_{-2}$ express the flicker and random walk noise levels, respectively.

Therefore, high optical absorption and responsivity \eqref{eq_RP} make the resonator more sensitive to laser intensity noise, highlighting a trade-off between responsivity and frequency fluctuations. This noise can be mitigated by selecting materials with low absorption in the targeted spectral range, or by operating the laser at its shot-noise limit $S_{I,\mathrm{sn}}(\lambda)$.

\section{Experimental results and discussion}
The experimental results focus on low-stress SiN resonators and are compared with the theoretical model (for details about the measurement procedures, see Materials and Methods).

\subsection{Strings}
\begin{figure*}
\includegraphics[width = 1\textwidth]{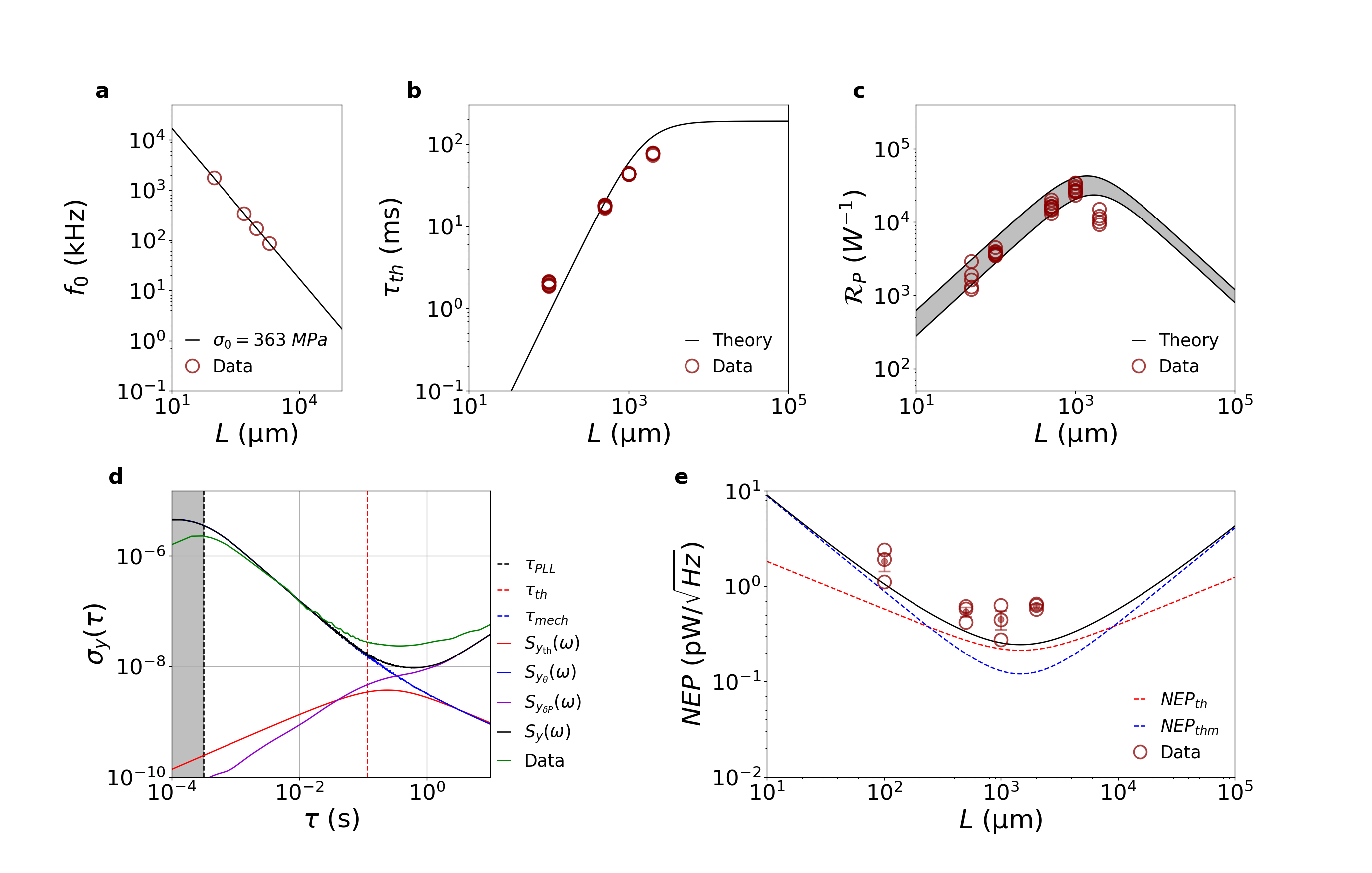}
\caption{\textbf{String design.} \textbf{a} Resonance frequency of $56$ nm thick, $5\ \mathrm{\mu m}$ wide string resonators of different lengths. From the measurements, a stress of 363~MPa is extracted. \textbf{b} Comparison between theoretical (black solid curve) and measured (dark red circles) thermal time constant $\tau_\mathrm{th}$ for the same set of strings. \textbf{c} Comparison between theoretical (solid curves) and measured (dark red circles) relative power responsivity. The error bar indicates the uncertainties in $\kappa$ ($2.7-4\ \mathrm{W/(m\ K)}$), $E$ ($200-300\ \mathrm{GPa}$). For these structures, $\alpha_\mathrm{th}=1\ \mathrm{ppm/K}$ has been measured. \textbf{d} Allan deviation measured for a $2\ \mathrm{mm}$ long string (green solid curve), driven at the onset of nonlinearity $z_\mathrm{r_c}$, with low-pass filter bandwidth $f_\mathrm{demod} = 2.5\ \mathrm{kHz}$, PLL bandwidth $f_\mathrm{pll} = 500\ \mathrm{Hz}$ and optical input power $P_0 = 6\ \mathrm{\mu W}$. The comparison with the theoretical model is also shown (black solid curve), together with the single contributions (see main text). \textbf{e} Comparison between the theoretical (black solid curve) and experimentally extracted (black circles) NEP for strings. The theoretical curve is composed of two different noise contributions: temperature (red dashed curve) and thermomechanical (blue dashed curve) fluctuations-induced fractional frequency noise. For each string's length, three different resonators were characterized in terms of NEP. Average and standard mean error for the data points are also shown for each length.}
\label{fig:String_plots}
\end{figure*}

Fig.~\ref{fig:String_plots}a shows the measured resonance frequency of SiN strings with varying length. The Q factor of these strings, essential for the theoretical calculations of the additive phase noise \eqref{eq_Sy_thm_closedloop}, has also been measured (for the data, see SI Section S3). 

Fig.~\ref{fig:String_plots}b displays the experimental thermal time constant (dark red circles) compared to theoretical predictions \eqref{eq_tauth}, showing excellent agreement. 

Fig.~\ref{fig:String_plots}c compares the theoretical model \eqref{eq_RP} (black solid lines) with the measured responsivity (dark red circles). The uncertainty band is defined by the uncertainty in material parameters $\kappa$ and $E$. 
All data points fall within the uncertainty band except for $L=2\ \mathrm{mm}$. This discrepancy stems from increased radiative losses caused by high probing optical power ($P_0=24-40\ \mathrm{\mu W}$). Higher incident powers lead to elevated temperatures at the string's center, increasing the radiative heat flux $\propto (T^4-T_0^4)$. This results in a nonlinear reduction of $\mathcal{R}_\mathrm{P}$, as well as a reduction in photothermal response time $\tau_\mathrm{th}$ (see Fig.~\ref{fig:String_plots}b).

The power responsivity can be enhanced by reducing the resonator's thickness $h$ and width $w$. On one hand, thinner strings will improve the thermal insulation, due to a reduction in cross-sectional area, as well as in emissivity \cite{Edalatpour2013}. On the other hand, narrower strings will reduce the hosting area for particle and molecule spectroscopy. Hence, the width choice is critical for photothermal sensing.

Fig.~\ref{fig:String_plots}d displays the Allan deviation (AD) for a string (green solid curve) \cite{Allan1966}. All the acquired AD have been compared with the theoretical model, accounting for the transfer functions \eqref{eq_Sy_thm_closedloop} of the PLL and SSO tracking schemes \cite{Zhang2023, Besic2023}. A good match is observed between measurements and theory (black solid curve) for integration times $\tau<0.1\ \mathrm{s}$, where the main noise source is additive in-phase (blue solid curve). For $\tau>0.1\ \mathrm{s}$, the data depart from the thermomechanical asymptote, with the presence of flicker frequency noise for $0.1\ \mathrm{s}<\tau< 1\ \mathrm{s}$, and frequency random walk for $\tau > 1\ s$, attributed to photothermal back-action (see below).

Fig.~\ref{fig:String_plots}e presents the resulting NEP values, evaluated at $\tau = \tau_\mathrm{th}$. For each length, three different chips have been analyzed (black circles). The results demonstrate strings' high photothermal sensitivity ($0.28-2.5\ \mathrm{pW/\sqrt{Hz}}$). The plot also displays the theoretical NEP (black solid curve), closely aligning with the measurements. The sensitivity is mainly limited by thermomechanical noise for almost all the lengths. The observed deviations must be attributed to the photothermal back-action.

\begin{figure}
    \begin{center}
    \centering
    \includegraphics[width = 0.5\textwidth]{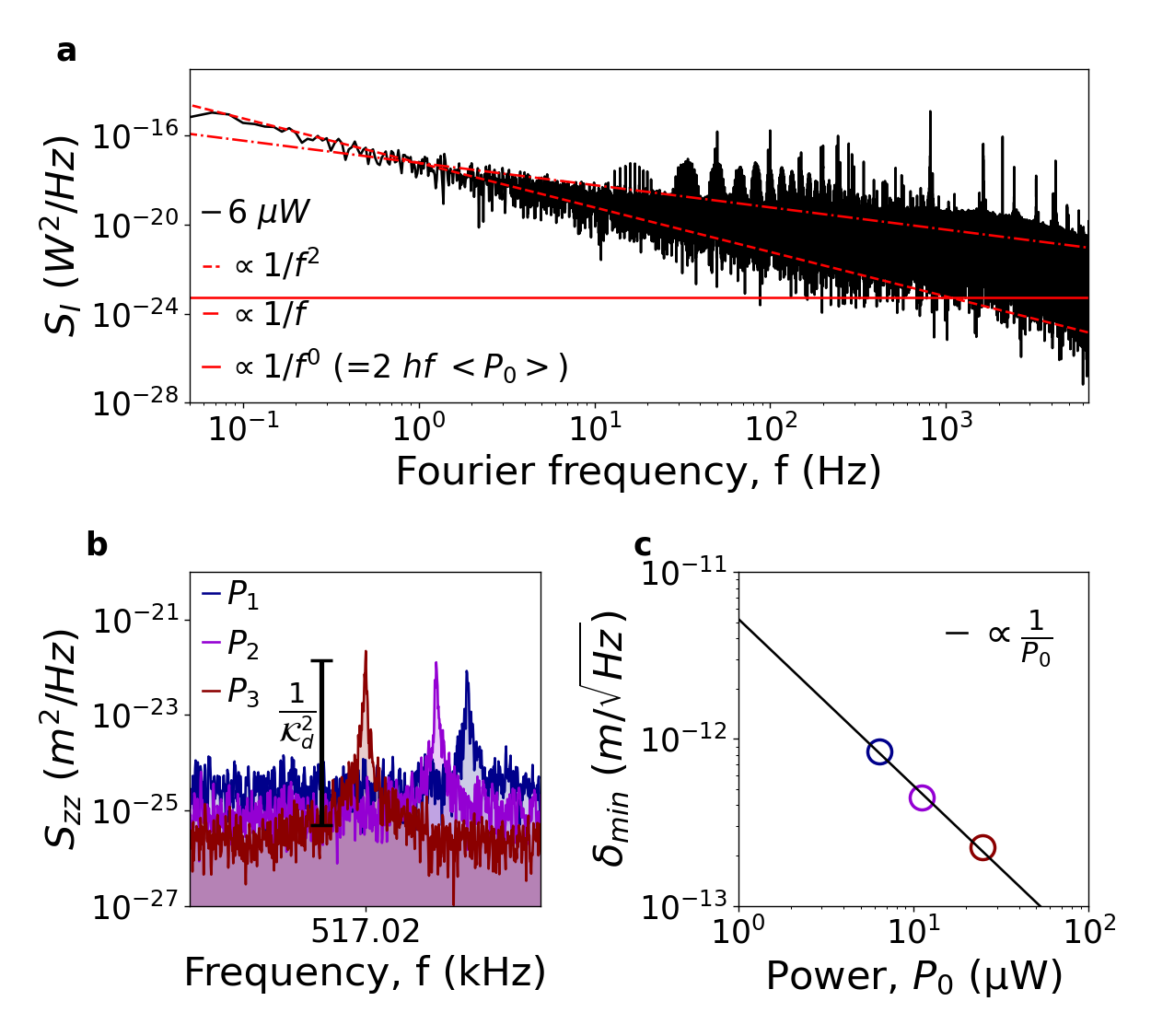}
    \caption{\textbf{Laser intensity fluctuations.} \textbf{a} Characterization of the intensity fluctuations for an average power $\braket{P_0}=6\ \mathrm{\mu W}$. The $f^{-2}$ and $f^{-1}$ noise contributions are shown (dashed red and dashed-dotted red lines, respectively). \textbf{b} Power spectral density of the thermomechanical noise for a drumhead resonator for different interferometer laser powers. \textbf{c}  Corresponding measured displacement sensitivity $\delta_\mathrm{min}$, in units [$\mathrm{m/\sqrt{Hz}}$]. It improves linearly with the laser power, with the effect of simultaneously introducing higher thermomechanical and laser power instability-induced frequency noise.} 
    \label{fig:Sxx_mbrn_P_comp}
    \end{center}
\end{figure}

The positive correlation between noise level measured for long integration times ($\tau>\tau_\mathrm{th}$) and power responsivity is indeed a clear sign of photothermal backaction \eqref{eq_Sypb}.  To investigate this further, the laser relative intensity noise $S_I(\omega, \lambda)$ has been characterized for all the optical powers employed in this study and $S_{y_\mathrm{\delta P}}(\omega, \lambda)$ evaluated. The results are displayed in Fig.~\ref{fig:String_plots}d with the purple solid curve, showing excellent agreement with the data for $\tau>0.1\ \mathrm{s}$. The observed flicker and random walk frequency noises are due to the intensity spectral distribution $S_I(\omega, \lambda)$, as clearly shown in Fig.\ref{fig:Sxx_mbrn_P_comp}a, far above the ultimate laser shot noise limit $S_{I,\mathrm{sn}}(\lambda)$.

Hence, photothermal back-action frequency noise imposes an upper limit on the probing power used for displacement transduction. On the one hand, high laser power improves the displacement sensitivity $\delta_\mathrm{min}$ [$\mathrm{m/\sqrt{Hz}}$], as shown in Fig.~\ref{fig:Sxx_mbrn_P_comp}b$\&$c \cite{Wagner1987}, reducing the detection coefficient $\mathcal{K}_d$. On the other hand, such a signal enhancement saturates at higher optical power due to the induced frequency noise \cite{Sadeghi2020}, with any low-frequency intensity noise, such as mode hopping \cite{Isenor1967}, directly impacting the resonator stability \cite{Maddaloni2013}.

Fig.~\ref{fig:Sxx_mbrn_P_comp}c further shows that the displacement sensitivity is here inversely proportional to the optical power $P_0$, indicating that the laser noise has a classical (detector and technical noise) and not quantum shot noise origin \cite{Lawall2000}. Among the different approaches to mitigate laser classical noise, feedback intensity noise squashing could offer a simple way to push the laser to its shot-noise limit \cite{Dumont2023}.

\subsection{Drumheads}
\begin{figure*}[t]
\includegraphics[width = 1\textwidth]{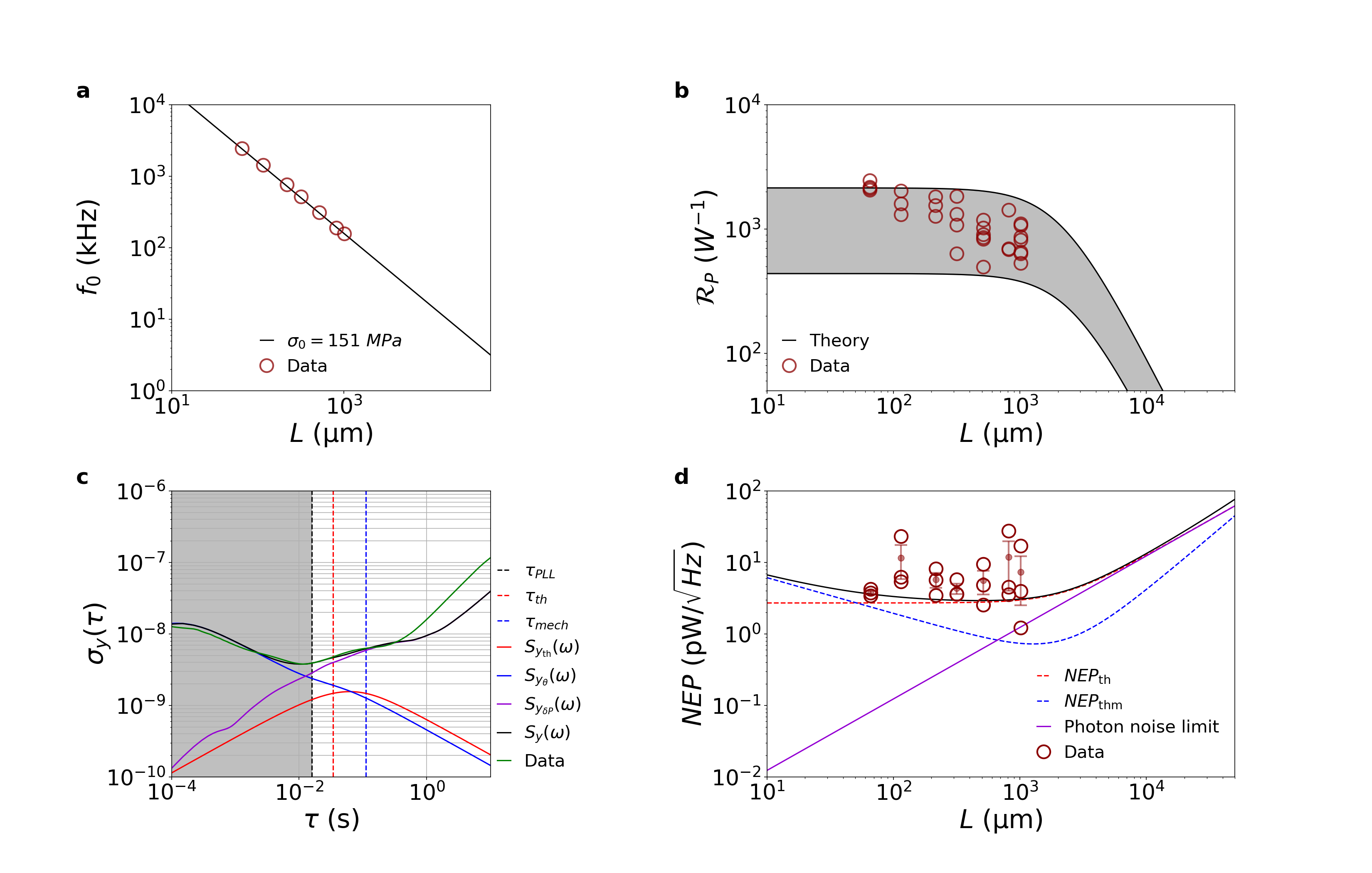}
\caption{\textbf{Drumhead design.} \textbf{a} Resonance frequency for $50$ nm thick square drumhead resonator of different side length. From the measurements, a stress of $150$ MPa is extracted. \textbf{b} Comparison between theoretical (solid curves) and measured (dark red unfilled dots) relative power responsivity. The error bar indicates the uncertainties in $\kappa$ ($2.7-4$ W/(m\ K)), $E$ ($200-300$ GPa), and $\alpha_\mathrm{th} (1-2.2\ \mathrm{ppm/K})$. \textbf{c} Allan deviation measured for a $1\ \mathrm{mm^2}$ square membrane (green solid curve), driven at the onset of nonlinearity $z_\mathrm{r_c}$, with low-pass filter bandwidth $f_\mathrm{demod} = 2.5$ kHz, PLL bandwidth $f_\mathrm{pll} = 10$ Hz and optical input power $P_0 = 6\ \mathrm{\mu W}$. The comparison with the theoretical model is also shown (black solid curve), together with the single contributions (see main text). \textbf{d} Comparison between the theoretical (black solid curve) and experimentally extracted (black unfilled dots) NEP for membranes. Temperature (red dashed curve) and thermomechanical (blue dashed curve) fluctuations-induced fractional frequency noise are also shown, together with the single photon noise limited NEP. For each membrane's length, three different resonators were characterized in terms of NEP.}
\label{fig:Mbrn_plots}
\end{figure*}

Fig.~\ref{fig:Mbrn_plots}a show the resonance frequency corresponding to the drumheads characterized experimentally (for the measured Q, see SI Section S3).
Experimental results concerning the thermal time constant are not presented here, as the photothermal response time of SiN drumheads has been already experimentally characterized elsewhere \cite{Piller2020, Snell2022}.

Fig.~\ref{fig:Mbrn_plots}b compares the theoretical predictions \eqref{eq_RP} (solid curves) with the experimental data (dark red circles) for the relative power responsivity. The uncertainty band, defined by uncertainties in $\kappa$, $E$, and $\alpha_\mathrm{th}$, encompasses all the experimental points, indicating a strong agreement between theory and experiments.

Fig.~\ref{fig:Mbrn_plots}c illustrates the AD for a drumhead. In detail, two regimes can be recognized: i) for different integration times $\tau<0.01\ \mathrm{s}$, the AD is limited by additive phase noise $S_{y_{\theta}} 
(\omega)$ (blue solid curve); ii) $\tau>0.01\ \mathrm{s}$, the noise is dominated by photothermal backaction $S_{y_\mathrm{\delta P}}(\omega)$. Notably, in the absence of photothermal backaction, temperature fluctuation frequency noise would dominate. This condition, where a mechanical resonator interacts with the environment at the single shot noise level, is of significant interest for micromechanical thermal detectors  \cite{Kruse2001a, rogalski2019infrared, Piller2023, Zhang2024}.

Fig. \ref{fig:Mbrn_plots}d presents the experimental NEP evaluated at $\tau=\tau_\mathrm{th}$, alongside the theoretical sensitivity (black solid curve), closely aligning to each other. The experimental results of $1-20\ \mathrm{pW/\sqrt{Hz}}$ are one order of magnitude lower than previously characterized, electrodynamically transduced drumhead resonators \cite{Piller2023}, showing the outstanding performances of pristine SiN structures over integrated nanoelectromechanical systems (NEMS), where electrodes are an important part of the design \cite{Chien2020a, Chien2020b, Piller2023}. The use of pure SiN for photothermal sensing applications is enabled by noninvasive transduction approaches, such as interferometry. In particular, pure optical transduction offers two key advantages: i) the absence of metal traces increases the thermal insulation, improving the responsivity \eqref{eq_RP}; ii), the sensor is not limited by Johnson noise, which usually degrades the frequency stability \eqref{eq_Sy} of a vast group of NEMS resonators \cite{Piller2023}. 
Conversely, bare SiN drumheads are mainly affected by temperature fluctuations noise (dark red dashed curve), as shown for $L>50\ \mathrm{\mu m}$. Moreover, as the resonator enters the radiation-limited regime, thermal photon shot noise becomes dominant (dark violet solid curve) \cite{Kruse2001a, Snell2022}.

\subsection{Trampolines}
The experimental analysis has been carried out for trampoline resonators with central pads designed using a Bezier profile \cite{Chien2020b, Pluchar2020, Vicarelli2022, Piller2023, Land2024}.

\begin{figure*}[t]
\includegraphics[width = 1\textwidth]{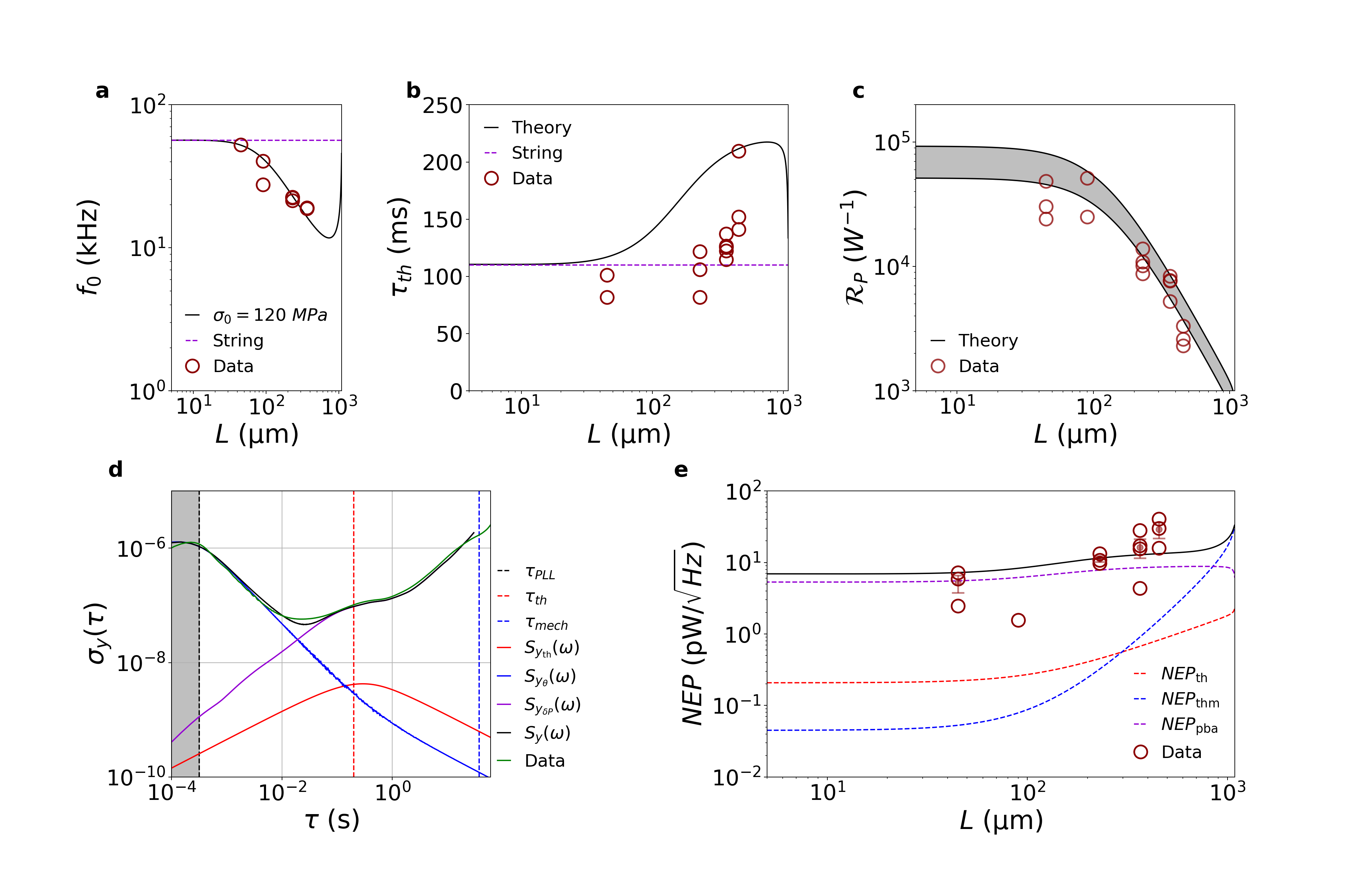}
\caption{\textbf{Trampoline design.} \textbf{a} Resonance frequency for $50$ nm thick trampoline resonators of different central area side lengths. The window side length is fixed to $L_\mathrm{w}\approx 1$ mm, while the tether width is to $w = 5 \mathrm{\mu m}$. From the measurements, a stress of $120\ \mathrm{MPa}$ is extracted. \textbf{b} Comparison between theoretical (black solid curve) and experimental thermal time constant $\tau_{th}$. \textbf{c} Comparison between theoretical (solid curves) and measured (dark red circles) relative power responsivity. The error bar indicates the uncertainties in $\kappa$ ($2.7-4\ \mathrm{W/(m\ K)}$), $E$ ($200-300\ \mathrm{GPa}$), and $\alpha_\mathrm{th}\ (1-2.2\ \mathrm{ppm/K}$). \textbf{d} Allan deviation measured for a $230^2\ \mathrm{\mu m}^2$ central area trampoline (green solid curve), driven at the onset of nonlinearities $z_\mathrm{r_c}$, with low-pass filter bandwidth $f_\mathrm{demod} = 2.5\ \mathrm{kHz}$, PLL bandwidth $f_\mathrm{pll} = 500\ \mathrm{Hz}$ and optical input power $P_0 = 11\ \mathrm{\mu W}$. The comparison with the theoretical model is also shown (black solid curve), together with the single contributions (see main text). \textbf{e} Comparison between the theoretical (black solid curve) and experimentally extracted (dark red circles) NEP. For each trampoline's central length, three different resonators were characterized in terms of NEP.}
\label{fig:Trmp_plots}
\end{figure*}

Fig.~\ref{fig:Trmp_plots}a presents the resonance frequency as a function of the central area $L^2$ (for the Q measurements, see SI Section S3). For small areas ($L^2<50^2\ \mathrm{\mu m^2}$), $\omega_0$ can be approximated with that of a string \cite{Schmid2023}. In the intermediate range ($50^2\ \mathrm{\mu m}^2<L^2<500^2\ \mathrm{\mu m}^2$) the effective mass $m_\mathrm{eff}$ grows faster  ($\propto L^2$) than the tethers' effective stiffness $k_\mathrm{eff}$ ($\propto L^{\zeta}$, with $\zeta<2$), leading to a reduction in resonance frequency $\omega_0$. For larger areas ($L^2>500\ \mathrm{\mu m}^2$) $k_\mathrm{eff}$ increases more rapidly than $m_\mathrm{eff}$ ($\zeta>2$), causing $\omega_0$ to rise beyond the string value (see SI Section S1).

Fig.~\ref{fig:Trmp_plots}b compares the theoretical thermal response time (black solid curve) with the experimental measurements (dark red circles). Discrepancies between the model and experimental data might be due to variations in specific heat capacity and mass density from the values used in the model. Nevertheless, a positive correlation between $\tau_\mathrm{th}$ and $L$ is evident. This slow thermal response significantly impacts the frequency noise in the experimental setup employed here.

Fig.~\ref{fig:Trmp_plots}c shows the comparison between the theoretical and measured power responsivity, exhibiting excellent agreement. As for the other designs, the shaded band represents uncertainties in $\kappa$, $E$, and $\alpha_\mathrm{th}$.

Fig.~\ref{fig:Trmp_plots}d shows the AD for a trampoline. Also here, two regimes can be recognized: an additive phase noise-limited region for integration times $\tau<0.02\ \mathrm{s}$, and a fully photothermal back-action frequency noise-dominated region for $\tau>0.02\ \mathrm{s}$. The sum of all the contributions (black solid curve) matches perfectly the experimental data (green solid). It is worth noting that $\tau_\mathrm{th}$ lies far in the photothermal back-action dominated region (red dashed vertical line), meaning that, during the time the resonator takes to reach a new thermal equilibrium, e.g. upon energy relaxation by a molecule, intensity fluctuations of the probing laser increase the frequency noise.
Conversely, with a shot-noise limited laser, the temperature fluctuation frequency noise would dominate for $\tau>\tau_\mathrm{th}$.   
 
Fig.~\ref{fig:Trmp_plots}e displays the experimental sensitivities evaluated at $\tau=\tau_\mathrm{th}$ (dark red circles) compared with the theoretical calculations (blue and red dashed curves). The plot reveals that the photothermal back-action (dark violet dashed curve) has degraded the sensitivity by one order of magnitude compared to the theoretical expectations. Moreover, this effect is much more pronounced for this design than for the others. Indeed, the slow thermal response time of trampolines makes them more sensitive to  the laser relative intensity noise \eqref{eq_Si}, where flicker and random walk noise are present, worsening the corresponding sensitivity
\begin{equation}
NEP_\mathrm{pba}=\alpha \sqrt{h_0 + 2\pi \tau_\mathrm{th}h_{-1} + (2\pi \tau_{th})^2h_{-2}}.
\end{equation}
However, the data follow the theoretical trend, with the sensitivity worsening for increasingly larger central areas $L^2$. Similar to drumheads, temperature fluctuations represent the ultimate theoretical limit of the photothermal sensitivity in the absence of photothermal back-action.

\subsection{Comparison}
\begin{figure}[h]
    \begin{center}
    \centering
    \includegraphics[width = 0.5\textwidth]{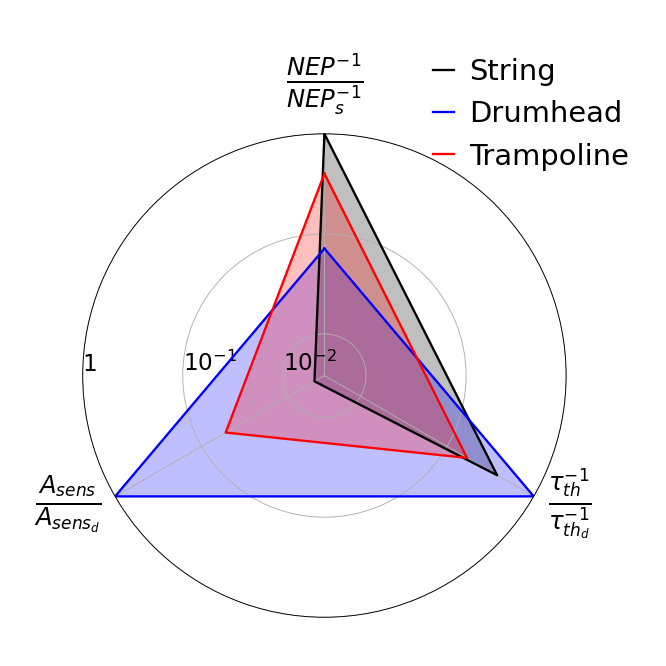}
    \caption{\textbf{Performance summary.} Radar chart of the nanomechanical photothermal performances. The chart accounts for the normalized NEP, thermal response time $\tau_\mathrm{th}$, and sensing area $A_\mathrm{sens}$. The highest value for each metric has been used for normalization, with the subscripts referring to the corresponding design (s, string; d, drumhead; t, trampoline). The string length, the membrane side length, and the trampoline window side length are all $1\ \mathrm{mm}$ long. The trampoline has a central area $L^2=230^2\ \mathrm{\mu m}^2$. All the structures are assumed to be $50\ \mathrm{nm}$ thick.}
    \label{fig:Radar}
    \end{center}
\end{figure}

To sum up, a theoretical comparison among the three resonator designs of comparable dimensions is illustrated in the radar chart shown in Fig.~\ref{fig:Radar}. The metrics used for this comparison are the NEP, the thermal time constant $\tau_\mathrm{th}$,  and the sensing area $A_\mathrm{sens}$, each normalized to the best-performing value. 

The string demonstrates the highest photothermal sensitivity due to its superior thermal insulation, albeit with the smallest sensing area. It presents an intermediate thermal response time compared to the other geometries. The fundamental frequency noise limit for this design is likely dominated by thermomechanical phase noise. These features make strings an excellent workhorse for nanomechanical photothermal spectroscopy \cite{West2023}. Conversely, the drumhead exhibits the lowest sensitivity but offers the largest sensing area and the fastest thermal response. In particular, the combination of high speed and optimal sensitivity for this design makes drumheads ideal for applications requiring quick measurements. Furthermore, temperature fluctuations are expected to be the ultimate frequency noise limit. 

Drumheads are good candidates for scanning spectromicroscopy, as well as a promising platform for room-temperature IR/THz detection. In particular, in the case of single-photon noise-limited detection, the large sensing area $A_\mathrm{sens}$ could enable drumheads to achieve the room-temperature specific detectivity limit $D^*=\sqrt{A_{sens}}/NEP \approx 2\cdot10^{10}\ \mathrm{cm \sqrt{Hz}}/W$ \cite{rogalski2019infrared, Snell2022, Piller2023}. 

Trampolines present a compromise between the highly sensitive strings and the drumheads with a larger sensing area. As such, trampolines show intermediate values in terms of power sensitivity and sensing area. Its only drawback is the slow thermal response, which makes it more susceptible to photothermal back-action frequency noise than the other designs, as confirmed by experimental observations. Despite this, its high sensitivity makes this design a good candidate for photothermal spectroscopy. Moreover, temperature fluctuations are expected to be the ultimate limiting frequency noise, therefore making it a promising alternative for IR/THz thermal detection and a potential competitor for drumheads.

The present study has examined the three main resonator designs exploited so far in nanomechanical photothermal sensing. Nonetheless, new designs routinely employed in other fields of nanomechanics, e.g., in optomechanics, could be explored for photothermal sensing. For instance, phononic crystal (PnC) engineering could be easily integrated within the sensor, enhancing the resonator frequency stability, as well as its thermal properties. In particular, the use of PnC defect modes for sensing applications would boost the power responsivity due to the increased overlap between the photothermally induced temperature field and the mechanical mode volume, as already shown \cite{Sadeghi2020pnc}.

\section{Conclusions}
In summary, the comparative analysis conducted on three distinct resonator designs utilized in photothermal sensing—namely strings, drumheads, and trampolines—has elucidated the relationship between the resonator's photothermal sensitivity and its mechanical and thermal properties. 
Across all scenarios, the theoretical framework has shown remarkable consistency with both experimental data and FEM simulations, demonstrating how the resonance frequency photothermal response is governed by the resultant mean temperature rise. 
Overall, strings emerge as the most sensitive design, followed by trampolines and drumheads.
Conversely, drumheads exhibit the fastest thermal response, followed by strings and trampolines.
The analysis has also highlighted the critical role of photothermal back-action, particularly its impact on the trampolines' frequency fluctuations, due to their slowest thermal response. These findings not only clarify the relative performance of the resonator designs investigated but also establish a solid groundwork for the development of next-generation nanomechanical photothermal sensors. This study represents a significant step forward in the advancement of nanomechanical sensing technology, offering valuable insights for researchers seeking to harness the full potential of photothermal sensing in diverse applications.

\section{Materials and methods}
\subsection{FEM simulations}
FEM analysis (COMSOL Multiphysics, v5.5 and v6.1) has been carried out to determine the relative power responsivity $\mathcal{R}_\mathrm{P}(0)$ and the thermal time constant $\tau_\mathrm{th}$ of the resonators. 
For $\mathcal{R}_\mathrm{P}(0)$, the Solid Mechanics Physics has been employed in conjunction with Heat Transfer in Solids. First, a Static Prestress Study is performed to solve for the resonator stress field, accounting for the initial prestress $\sigma_0$ and the thermal stress components induced by laser heating. A Gaussian beam with beam waist $w_0$ is used as the light source. The static solutions obtained from this study serve as input parameters for the Eigenfrequency Study, where the fundamental eigenfrequency is computed. The procedure is repeated for different input powers $P_0$.
For the evaluation of $\tau_\mathrm{th}$, a time-dependent study is conducted with the beam impinging at the resonator center. The temperature of the resonator is solved for discrete points in time, and the thermal time constant is evaluated upon fitting with an exponential function.

\subsection{Experimental setup}
The resonators are operated in high-vacuum conditions ($p < 10^{-5}\ \mathrm{mbar}$) to minimize gas damping and thermal convection, in a custom-designed vacuum chamber equipped with a window for optical access to the chips. The resonators are actuated with a piezoelectric element placed beneath them. Their out-of-plane displacement is measured with a commercial laser-Doppler vibrometer (MSA 500, Polytec Gmbh), operated at 633 nm wavelength. The vibrometer's signal is sent to a lock-in amplifier (HF2LI, Zurich Instruments) equipped with a PLL module, or to a frequency counter implemented in a self-sustained oscillator (PHILL, Invisible-Light Labs GmbH) \cite{Besic2023}.
The relative power responsivity $\mathcal{R}_\mathrm{P}(0)$ is evaluated upon measurement of the thermomechanical noise spectrum peak for different input laser powers \cite{Sadeghi2020pnc}. The thermal time constant $\tau_\mathrm{th}$ is evaluated with the 90/10 method \cite{Piller2023}. For that, the resonator is driven at its resonance frequency with the PLL or SSO tracking scheme.

\subsection{Laser intensity fluctuations characterization}
The intensity of the probing laser has been recorded for 1 minute with a silicon photodiode (Thorlabs GmbH S120C, $1\ \mathrm{\mu s}$ response time) together with a digital power meter console (Thorlabs GmbH PM100D). The electrical signal is fed to the lock-in amplifier \cite{Sadeghi2020}. The recorded intensity signal is then converted into frequency fluctuations, accounting for the resonator's thermal response $H_\mathrm{th}(\omega)$, and the corresponding AD is evaluated (See SI Section S4).

\section{Acknowledgement}
The authors are thankful to Sophia Ewert and Patrick Meyer for the device fabrication. Furthermore, the authors thank Johannes Hiesberger, Hajrudin Besic and Niklas Luhmann for useful discussions. 
This work received funding from the Defense Advanced Research Projects Agency (DARPA) Optomechanical Thermal Imaging (OpTIm) Technical Area (TA) 1 Broad Agency Announcement (BAA), HR001122S0055. This work further received funding from the Novo Nordisk Foundation under project MASMONADE with project number NNF22OC0077964, and from the European Innovation Council under the European Union’s Horizon Europe Transition Open program (Grant agreement: 101058711-NEMILIES).

\section{Contributions}
K.K, and S.S. designed research; K.K., and S.E. performed experiments; K.K., and F.L. performed FEM simulations; K.K., F.L., P.M., R.G.W., and S.S. analyzed the data; K.K., F.L., R.G.W., and S.S. wrote the paper with inputs from all authors; S.S. supervised the project.

\section{Conflict of interests}
S.S. is co-founder of Invisible-Light Labs GmbH, which provided the PHILL electronics.

\section{Supplementary information}
The online version contains supplementary materials.

\bibliography{nep_biblio}

\end{document}


\title{Supplementary information: Comparative Analysis of Nanomechanical Resonators: Sensitivity, Response Time, and Practical Considerations in Photothermal Sensing}
\author{Kostas Kanellopulos}
\author{Friedrich Ladinig}
\author{Stefan Emminger}
\author{Paolo Martini}
\author{Robert G. West}
\author{Silvan Schmid}
\email[Correspondence email address: ]{silvan.schmid@tuwien.ac.at}
        \affiliation{Institute of Sensor and Actuator Systems, TU Wien, Gusshausstrasse 27-29, 1040 Vienna, Austria.}

\date{\today} 

\maketitle

\tableofcontents

\newpage

\section{Mechanics of the trampoline}
\begin{figure}[h!]
    \begin{center}
    \centering
    \includegraphics[width = 1\textwidth]{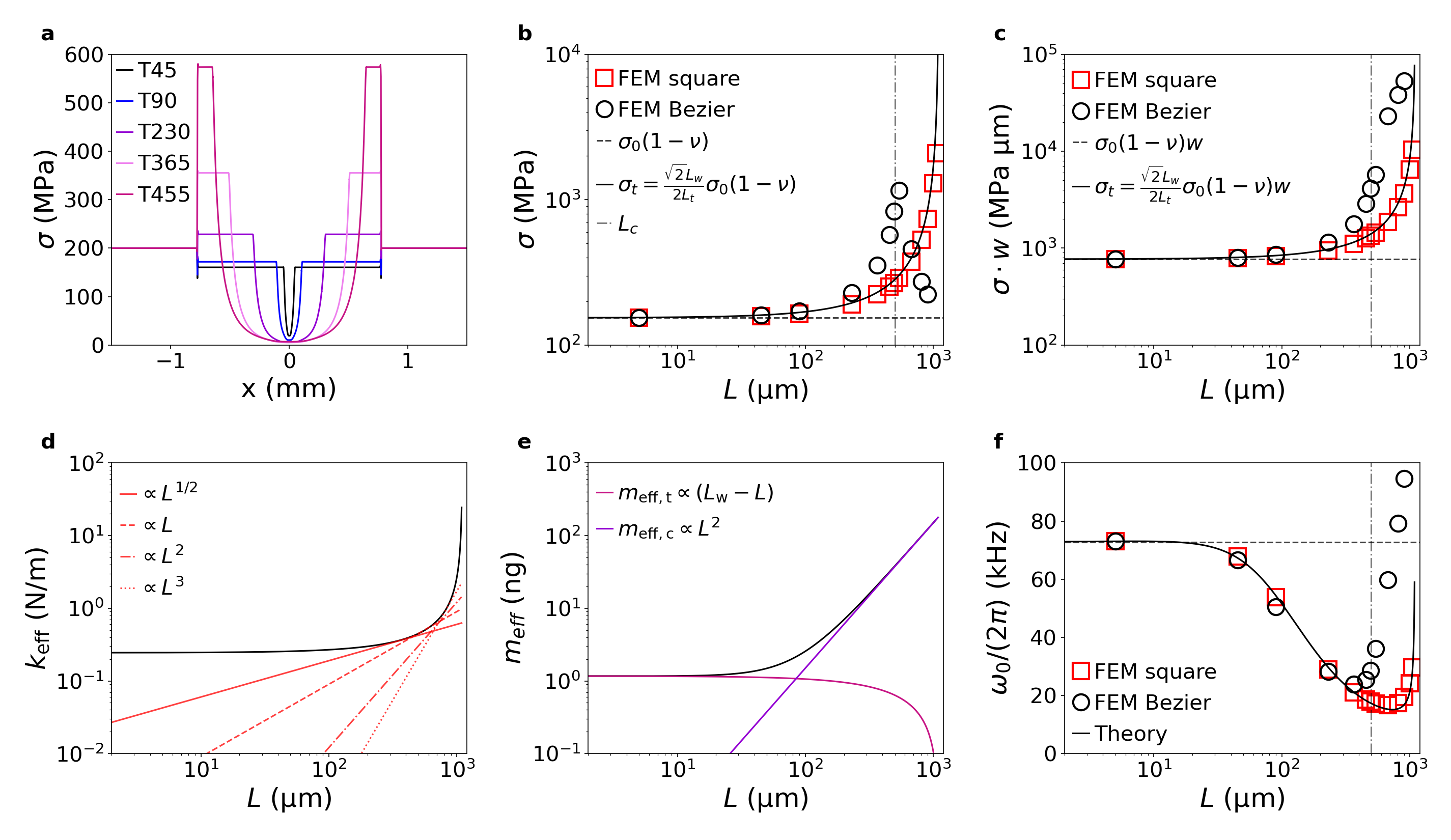}
    \caption{\textbf{Mechanics of the trampoline.}
    \textbf{a} X-cut stress profile in trampolines with a central pad with a Bezier curve profile.
    \textbf{b} Tethers stress as a function of the central pad side length $L$. Red squares: FEM simulations for a trampoline with a square design of the central pad. Black circles: FEM simulations for a trampoline with a Bezier profile design for the pad. Black curve: theory \eqref{eq_sigma_t}. Vertical dashed dotted line: critical central pad side length $L_c$. For $L>L_c$, the tether width at the clamping points for a Bezier trampoline increases with $L$, changing the boundary conditions relative to the square design.
    \textbf{c} Product tether's stress-width $\sigma_t\cdot w$ as a function of $L$. For $L>L_c$, the reduction in stress is compensated by the increase in width at the clamping points.
    \textbf{d} Square trampoline spring constant \eqref{eq_kteff} as a function of $L$. Displayed are also different power laws $L^{\zeta}$ for clarity. \textbf{d} Trampoline effective mass as a function of $L$. Pink curve: tethers' effective mass. Purple: central pad's effective mass.
    \textbf{f} Fundamental resonance frequency as a function of $L$.
    FEM and model parameters: $\rho = 3000\ \mathrm{kg/m^3}$, $E = 250\ \mathrm{GPa}$, $\sigma_0=200\ \mathrm{MPa}$, $\nu=0.23$, $h=50\ \mathrm{nm}$. For the square design, $w = 5\ \mathrm{\mu m}$ always; for the Bezier design, $w = 5\ \mathrm{\mu m}$ for $L\leq L_c$.}
    \label{fig:keff_meff_trmp}
    \end{center}
\end{figure}

A trampoline resonator can be modeled using a lumped element approach, where an effective mass $m_\mathrm{eff}$ is connected to a fixed frame (the square window of side length $L_\mathrm{w}$), via a spring of constant $k_\mathrm{eff}$ (representing the diagonal four tethers). Under the assumption of a tensile force $N$ applied on the unsuspended thin film of thickness $h$, and further considering that the resulting strain $\epsilon$ remains constant after the release process (since the distance between clamping points is unchanged), the balance of the tensile force can be expressed as \cite{Fedorov2019}
\begin{equation} \label{eq_N}
    \frac{N}{h E} = const = \epsilon(x) w(x) = \frac{\sigma(x)}{E} w(x).
\end{equation}
$w(x)$ denotes the local width of the geometry, function of the coordinate x along a cut-line. From Eq.~\eqref{eq_N}, it can be seen that the tethers concentrate higher stress $\sigma_\mathrm{t}$ than the central pad, due to a reduction in cross-section. This is clearly illustrated in Fig.~\ref{fig:keff_meff_trmp}a, where a cut-line along the x coordinate is shown for FEM simulated trampolines with a Bezier profile. Moreover, as the tethers shorten (for high $L$ values), this stress further increases. The FEM model includes also the chip, to better show the stress distribution. 
For clarity, FEM simulations have been also performed for the simplest trampoline geometry: a central square pad of area $L^2$ and effective mass $m_\mathrm{eff,c}$, connected to the frame via four tethers along its two diagonals, each of length $L_\mathrm{t}$ and effective mass $m_\mathrm{eff,t}$. For such a trampoline oscillating at its fundamental resonance frequency $\omega_0$, the effective spring constant $k_\mathrm{eff}$ can be modelled as that of a string of length $L_\mathrm{t}$, under a prestress $\sigma_0 (1-\nu)$, which is given by
\begin{equation} \label{eq_kteff}
    k_\mathrm{eff}(\sigma_0, L_\mathrm{t}) =\ \frac{\pi^2}{2} \frac{w h}{L_\mathrm{t}} \sigma_0 (1 - \nu)
\end{equation}
with $\sigma_0$ denoting the nominal stress of the unstructured thin film, and with the factor $(1-\nu)$ accounting for the transverse strain relaxation upon release. From Eq.~\eqref{eq_kteff}, it is possible to extract the stress concentrated at the tethers
\begin{equation} \label{eq_sigma_t}
    \sigma_\mathrm{t} =\ \frac{2}{\pi^2} \frac{1}{wh} k_\mathrm{eff} \frac{\sqrt{2}L_\mathrm{w}}{2} = 
    \frac{\sqrt{2} L_\mathrm{w}}{2 L_\mathrm{t}} \sigma_0 (1 - \nu).
\end{equation}
Hence, $\sigma_\mathrm{t}$ is directly proportional to the ratio of the trampoline diagonal length ($\sqrt{2} L_\mathrm{w}$) to the total length of the two parallel tethers ($2 L_\mathrm{t}$).
Fig.~\ref{fig:keff_meff_trmp}b displays this theoretical stress component \eqref{eq_sigma_t} as a function of the central pad side length $L$ (black curve), closely aligning with the FEM results (red squares). As expected, the stress increases with $L$. 
For comparison, the FEM results for the trampolines with a Bezier profile are also displayed (black circles). Below a critical side length ($L\leq L_c\approx 500\ \mathrm{\mu m}$), the stress at the tethers grows faster with $L$ than what observed for the square design. For $L>L_c$, the tethers' stress drops down, as expected from Eq.~\eqref{eq_N}. Indeed, their width increases with $L$ in this region, conversely to the square design case, for which $w$ is constant. This increase in $w$ compensates for the stress reduction, making the Bezier trampolines stiffer than the square design for $L>L_c$ (Fig.~\ref{fig:keff_meff_trmp}c), consistent with the FEM simulated fundamental resonance frequency (see below, Fig.~\ref{fig:keff_meff_trmp}f). Fig.~\ref{fig:keff_meff_trmp}c clearly illustrates this compensation with the product tethers' stress-width as a function of $L$.
Fig.~\ref{fig:keff_meff_trmp}d shows the corresponding effective spring constant $k_\mathrm{eff}$ \eqref{eq_kteff} as a function of $L$. Different power laws are displayed to illustrate the change in spring constant with central area growth. For $L^2<200^2\ \mathrm{\mu m^2}$, the stiffness matches the case of a simple string resonator, as here the trampoline is cross-string structure. For $L^2 > 200^2\ \mathrm{\mu m^2}$, the stiffness increases significantly, due to stress concentration at the tethers. 

From the modeshape of the trampoline's fundamental resonance, its effective mass $m_\mathrm{eff}$ can be written as
\begin{equation} \label{eq_mteff}
    m_\mathrm{eff} =\ m_\mathrm{c,eff}\ +\ m_\mathrm{t,eff} =\ \rho\ h\ \left (L^2\ +\ \frac{4\ w\ L_\mathrm{t}}{2}\right ),
\end{equation}
with $\rho$ denoting the resonator mass density. The tether's effective mass $m_\mathrm{t,eff}$ is the same as for the string, $m_\mathrm{s,eff}=0.5\ m_0$. For the central pad, $m_\mathrm{c,eff}$, its full inertial mass is accounted for as the entire pad is being displaced for the fundamental mode. Fig.~\ref{fig:keff_meff_trmp}e displays $m_\mathrm{eff}$ as a function of the central pad side length $L$. Two regimes can be seen: for $L^2<100^2\ \mathrm{\mu m^2}$, the mass remains almost constant, as the reduction in tether length is counterbalanced by the growth of the central pad; for $L^2> 100^2\ \mathrm{\mu m^2}$, the central pad fully defines the effective mass, growing here as $m_\mathrm{eff}\propto L^2$. It is worth noting that, in the range $100\ \mathrm{\mu m}<L<500\ \mathrm{\mu m}$, the effective mass grows faster than the spring constant, reducing the overall resonance frequency in this region (Fig.~\ref{fig:keff_meff_trmp}f).
Fig.~\ref{fig:keff_meff_trmp}f compares the FEM results for square and Bezier trampolines with the theoretical predictions for the fundamental resonance frequency $\omega_0 = \sqrt{k_\mathrm{eff} / m_\mathrm{eff}}$, showing excellent agreement. $\omega_0$ increases faster with $L$ for the Bezier design, due to an overall increase in tethers' stiffness, as shonw above.

Having a closed expression for the trampoline's fundamental resonance frequency allows for the extraction the relative temperature responsivity $\mathcal{R}_T$.
Whenever the resonator experiences a mean temperature rise $\braket{\Delta T}$, the effective stiffness is reduced by the built-in thermal stress along the tethers
\begin{equation} \label{eq_kteff_demon}
    k_{\mathrm{eff}}(T) =\frac{\pi^2}{2} \frac{w h}{L_{\mathrm{t}}} (1-\nu) \left [\sigma_0 - \alpha_{\mathrm{th}} E \braket{\Delta T} \right] = 
    k_\mathrm{eff}(\sigma_0, L_\mathrm{t}) \left[1-\frac{\alpha_\mathrm{th} E}{\sigma_0} \braket{\Delta T}\right].
\end{equation}
Substituing Eq.~\eqref{eq_kteff_demon} into the resonance frequency equation gives
\begin{equation}\label{eq_omega0_trmp_T}
    \omega_0(T) = \sqrt{\frac{k_\mathrm{eff}(T)}{m_\mathrm{eff}}} =
    \sqrt{\frac{k_\mathrm{eff}(\sigma_0, L_\mathrm{t})}{m_\mathrm{eff}}}\sqrt{1 - \frac{\alpha_\mathrm{th} E}{\sigma_0}\braket{\Delta T}} \approx
    \omega_0(0) \left[1 - \frac{\alpha_\mathrm{th} E}{2 \sigma_0} \braket{\Delta T}\right]
\end{equation}
Therefore, the temperature responsivity $\mathcal{R}_T$ for a trampoline is recovered (see Eq.~12 in the main text).

\section{Mean temperature framework (MTF)}

\begin{figure}[h]
    \begin{center}
    \centering
    \includegraphics[width = 0.75\textwidth]{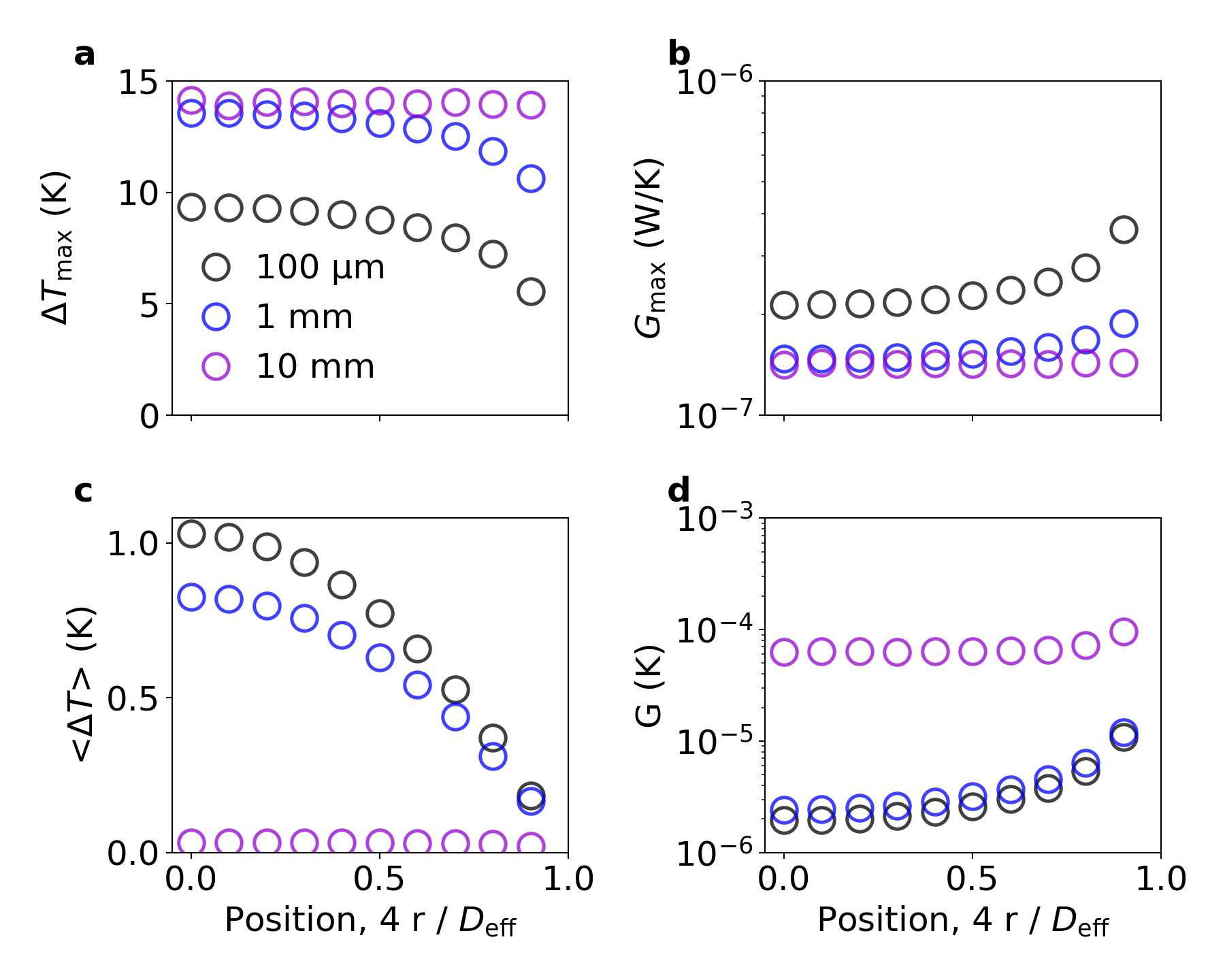}
    \caption{\textbf{Mean temperature framework.} FEM simulations of drumhead resonators heated by a laser of power $P_0 = 10\ \mathrm{\mu m}$ and beam waist of $1\ \mathrm{\mu m}$, for different resonator size: $100\ \mathrm{\mu m}$, black empty dots; $1$ mm, blue; $10$ mm, dark violet. \textbf{a.} Maximum temperature rise $\Delta T_\mathrm{max}$ as a function of the laser position. \textbf{b.} Corresponding thermal conductance $G_\mathrm{max} = P_0 / \Delta T_\mathrm{max}$. \textbf{c.} Mean temperature $\braket{\Delta T}$ for the same laser conditions. \textbf{d.} Corresponding thermal conductance $G=P_0/\braket{\Delta T}$.}
    \label{fig:MTF_mbrn}
    \end{center}
\end{figure}

The resonance frequency is a global property of a resonator, depending on its material and geometry. Consequently, variations in resonance frequency are expected to be dictated by the mean temperature changes $\braket{\Delta T}$, rather than the local variations $\Delta T$. This has been clearly shown in Fig.~2h. To further underline this point, FEM simulations for circular drumheads of different sizes ($L = 100\ \mathrm{\mu m}$, black; $1$ mm, blue; $10$ mm, dark violet) are performed with a tightly focused heating laser, for different laser positions (Fig.~\ref{fig:MTF_mbrn}). 
Fig.~\ref{fig:MTF_mbrn}a$\&$b display the FEM results for the maximum temperature rise $\Delta T_\mathrm{max}$ at thermal equilibrium, and the corresponding thermal conductance $G_\mathrm{max}=P_0/\Delta T_\mathrm{max}$, respectively. For the concentric case ($r = 0$), $\Delta T_\mathrm{max}$ increases for larger drumheads, making the largest resonator the most thermal insulated. Fig.~\ref{fig:MTF_mbrn}c$\&$d display the mean temperature rise $\braket{\Delta T}$ and corresponding thermal conductance $G=P_0/\braket{\Delta T}$ for the same FEM simulations, respectively. Again, looking only at the concentric case, these results show a completely opposite trend than what it is shown in Fig.~\ref{fig:MTF_mbrn}a$\&$b: the mean temperature rise at thermal equilibrium reduces with larger resonators,  leading to an increase in conductance with the drumhead's lateral dimensions. Furthermore, the simulated orders of magnitudes of $G$ are consistent with the power responsivity analysis discussed in the main text (see Fig.~2g). 
Hence, the mean temperature framework (MTF) must be employed to accurately describe the photothermal response of the resonator. Within this framework, the shape $s_f$ and the $\beta$ factors must be introduced. These quantities redefine the thermal losses due to heat conduction in the MTF as
\begin{equation}\label{eq_Gcond_mtf}
    G_\mathrm{cond} = \frac{s_f(\mathbf{r},L,w_0)}{\beta(\mathbf{r}, L, w_0)} \kappa, 
\end{equation}
with $\mathbf{r}=(x,y)$ or $(r,\theta)$, $L$, and $w_0$ denoting the heat source position vector, the resonator characteristic length, and the heat source size, respectively.

\subsection{String}
\subsubsection{Shape and $\beta$ factors}
\begin{figure}
    \begin{center}
    \centering
    \includegraphics[width = 0.75\textwidth]{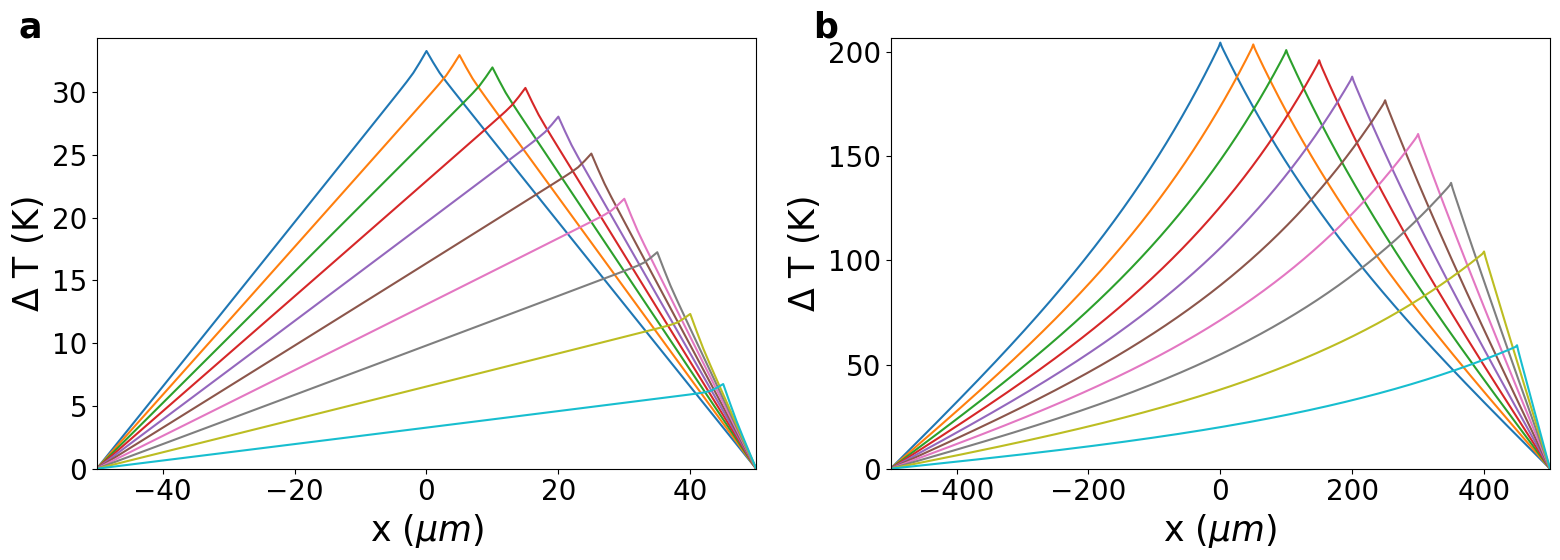}
    \caption{\textbf{String temperature profile.} \textbf{a} FEM simulated 1D temperature profiles of a $100\ \mathrm{\mu m}$ long string resonator, for different positions of the heat source.  \textbf{b} FEM simulated 1D temperature profiles for a $1$ mm long string. Laser parameters: input optical power $P_0 = 200\ \mathrm{\mu W}$, beam waist $w_0 = 1\ \mathrm{\mu m}$. FEM parameters: $\rho = 3000\ \mathrm{kg/m^3}$, $c_p=700\ \mathrm{J/(kg\ K)}$, $\kappa = 3\ \mathrm{W/(m\cdot K)}$, $E = 250\ \mathrm{GPa}$, $\sigma_0=200\ \mathrm{MPa}$, $\nu=0.23$, $\alpha_\mathrm{th} = 2.2\ \mathrm{ppm/K}$, $\epsilon_\mathrm{rad} = 0.05$, $\alpha_\mathrm{abs}=0.5\ \%$, $w = 5\ \mathrm{\mu m}$, $h=50\ nm$.}
    \label{fig:SF_strg}
    \end{center}
\end{figure}

The string design represents the simplest geometry from a thermal transport standpoint. Considering a tightly focused laser as heat source, impinging with power $P_0$ in a position $x$ along the length of the resonator in thermal equilibrium with it, the Fourier law's gives \cite{Schmid2023}
\begin{equation} \label{eq_Fourier_string}
    P_0 = \frac{4 w h}{L} \kappa \Delta T_\mathrm{max} = s_f(x,L,w_0) \kappa \Delta T_\mathrm{max},
\end{equation}
with $w$, $h$, and $L$ being the resonator width, thickness and length, respectively. $\Delta T_\mathrm{max}=T_\mathrm{max} - T_0$ denotes the peak temperature rise with respect to the frame temperature $T_0$, occurring at the heat source position. For a string resonator, such as those analyzed in the main text, a linear temperature profile is the solution of the heat diffusion equation in steady-state for short and intermediate length ($L\leq2\ \mathrm{mm}$) 
\begin{equation}\label{eq_Tprofile_strg}
    \Delta T(x) = \Delta T_\mathrm{max} - \frac{2}{L} \Delta T_\mathrm{max} |x|,\ \ \ \mathrm{for}\ -\frac{L}{2}\leq x\leq\frac{L}{2},
\end{equation}
as shown in Fig.~\ref{fig:SF_strg}. The 1D temperature profiles for $100\ \mathrm{\mu m}$ (left) and $1$ mm (right) long strings have been obtained using FEM for different heating laser positions. For both strings, all profiles are linear. In the $1$ mm long string, the thermal radiation plays a more significant role than in the $100\ \mathrm{\mu m}$ long one, causing the profile to deviate slightly from a purely linear trend. Nonetheless, as long as $\Delta T$ can be treated as a linear function of the position $\mathbf{r}$, even in the presence of thermal radiation losses, the following geometrical relation holds true
\begin{equation}\label{eq_T_string}
    \braket{\Delta T}=\frac{1}{L} \int_{-L/2}^{L/2} \Delta T(x) dx = \frac{1}{L} \frac{L \Delta T_\mathrm{max}}{2}=\frac{\Delta T_\mathrm{max}}{2},
\end{equation}
yielding the $\beta$ factor for a string resonator as $\beta = \frac{1}{2}$.

\subsubsection{Heat localization}
\begin{figure}[h]
    \begin{center}
    \centering
    \includegraphics[width = 0.5\textwidth]{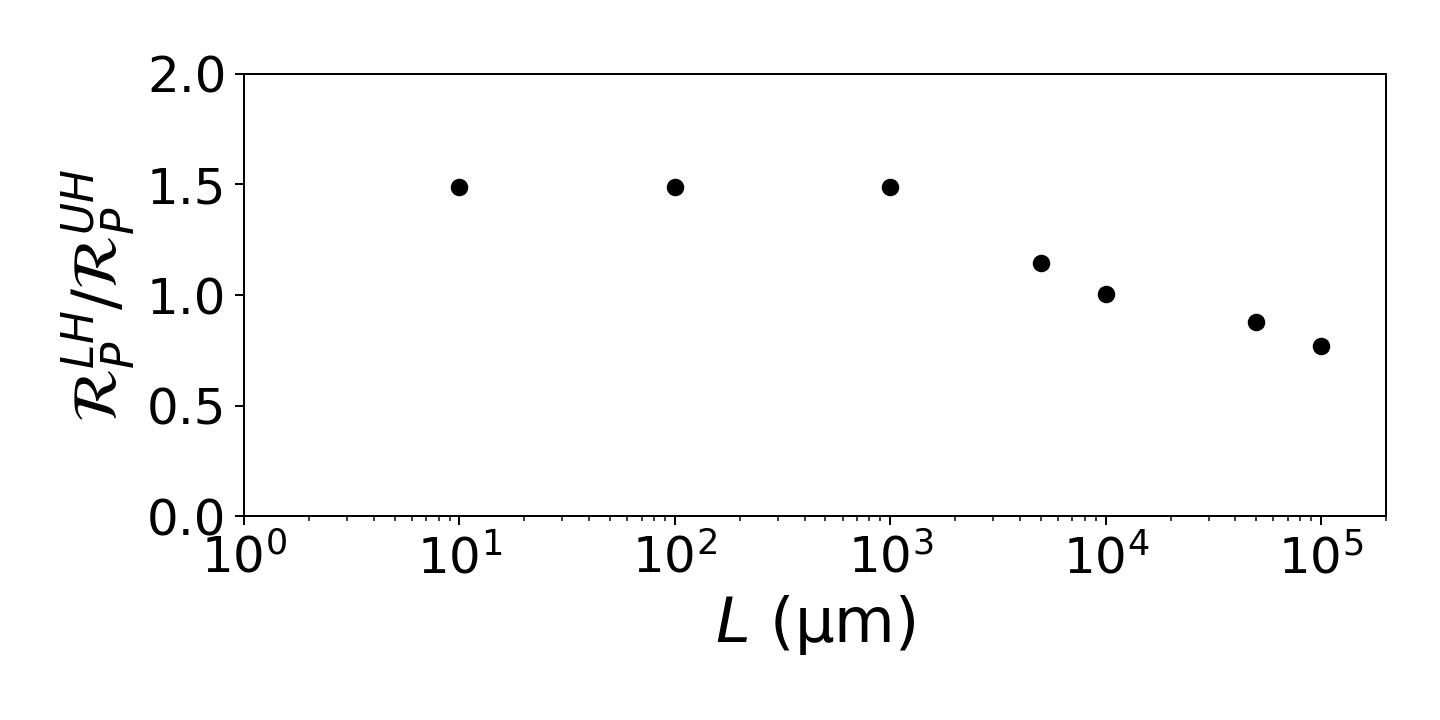}
    \caption{\textbf{Heat localization in strings.} Ratio between power responsivity for localized (LH) and uniform (UH) heating conditions. FEM parameters: $\rho = 3000\ \mathrm{kg/m^3}$, $c_p=700\ \mathrm{J/(kg\ K)}$, $\kappa = 3\ \mathrm{W/(m\cdot K)}$, $E = 250\ \mathrm{GPa}$, $\sigma_0=200\ \mathrm{MPa}$, $\nu=0.23$, $\alpha_\mathrm{th} = 2.2\ \mathrm{ppm/K}$, $\epsilon_\mathrm{rad} = 0.05$, $\alpha_\mathrm{abs}=0.5\ \%$, $w = 5\ \mathrm{\mu m}$, $h=50\ nm$, $P_0=10\ \mathrm{\mu W}$.}
    \label{fig:ratio_loc_strg}
    \end{center}
\end{figure}

Nanomechanical photothermal sensing can be performed with tightly focused as well as uniformly distributed heat sources / beam diameters. Greater (lesser) localization of the heating yields higher (lower) temperature rises $\braket{\Delta T}$. For a string resonator, two types of FEM simulations have been carried out: i) local heating (LH) with a point heat source at the string's center; ii) uniform heating (UH) with the upper surface defined as the heating source. No Gaussian beam lasers are used here, since part of the total input power would be lost in the uniform heating condition, perpendicular to the string length. The ratio $\mathcal{R}_P^{LH}/\mathcal{R}_P^{UH}$ between the LH and UH power responsivity is plotted as a function of the string length in Fig.~\ref{fig:ratio_loc_strg}.
For $L\leq1$ mm, this ratio is constant at 1.5, indicating that localized heating provides a $1.5\times$ improvement in power responsivity compared to uniform heating. For longer strings ($L>1$ mm), the highly localized optical power at the center increases the radiation losses $\propto (T^4 - T_0^4)$, worsening the responsivity improvement.

For uniform illumination, the point along the string length will contribute to the heat dissipation. Integrating \eqref{eq_Gcond_mtf} for a string, for a concentric source gives
\begin{equation}
    G_\mathrm{cond}= \left( \frac{1}{\kappa}\frac{1}{L}\int_0^L{\frac{\beta}{s_\mathrm{f}(x)}}\text{d}x\right)^{-1} = 12 \frac{w h}{L}\kappa.
\end{equation}
The overall conductance for a uniformly heated string is given by
\begin{equation}
    G = 12 \frac{w h}{L}\kappa  + 8 w L \epsilon_\mathrm{rad} \sigma_\mathrm{SB} T_0^3
\end{equation}

\subsection{Drumhead}
\subsubsection{Shape and $\beta$ factors}
For simplicity, a circular membrane of diameter $D_\mathrm{eff} = 2 L/\sqrt{\pi}$ is considered here, heated by a laser source of beam waist $w_0$ centered at position $(r, \theta)$ relative to the membrane center (Fig.~2e). Given that $h<<D_\mathrm{eff}$, no thermal gradient along the resonator thickness is present, consistent with the eccentric shell scenario \cite{Bergman2017}. 
\begin{figure}[h!]
    \begin{center}
    \centering
    \includegraphics[width = 0.75\textwidth]{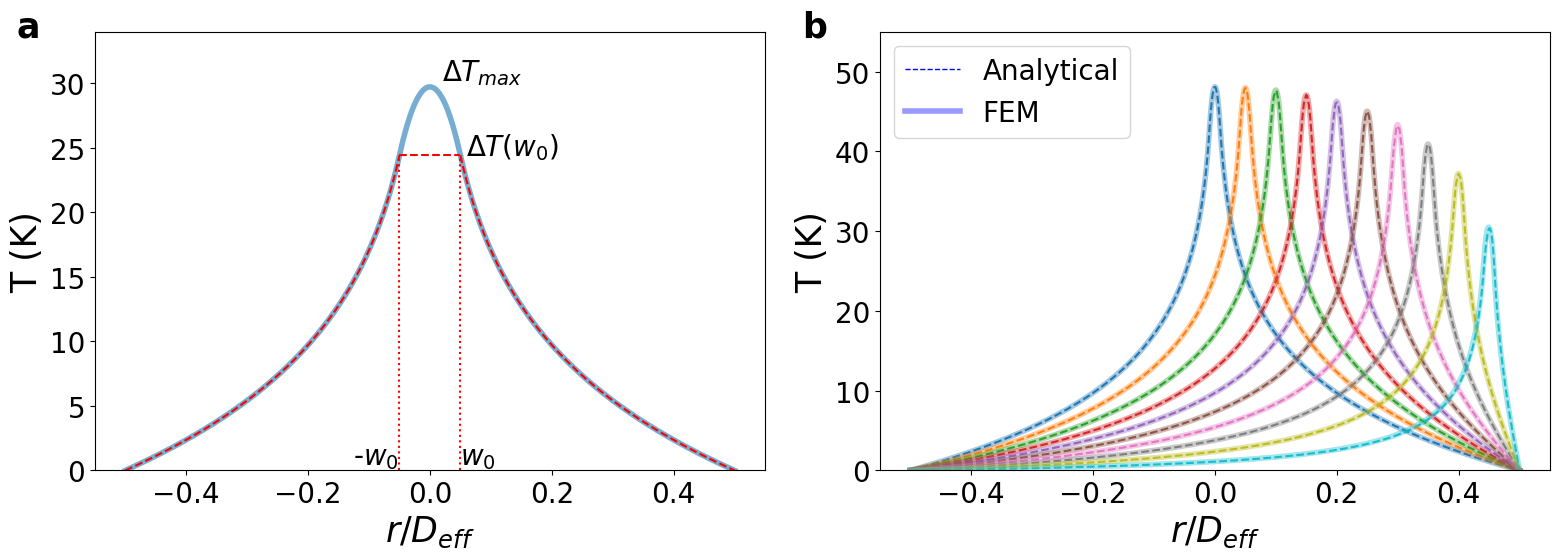}
    \caption{\textbf{Membranes temperature profile.} \textbf{a} 1D Temperature profile of circular membrane of diameter $D_\mathrm{eff}$, heated in the center by a top-hat disk beam of diameter $d$ (light blue solid curve). For comparison, the temperature distribution in the case of a eccentric cylinder, uniformly heated is shown (red dashed curve). \textbf{b} Comparison between FEM (solid curves) and analytical (dashed curves) temperature profiles, obtained for a localized heat source of input power $P_0=10\ \mathrm{\mu W}$ and beam waist $w_0=1\ \mathrm{\mu m}$, moving along a radial cut-line. FEM parameters: $\rho = 3000\ \mathrm{kg/m^3}$, $c_p=700\ \mathrm{J/(kg\ K)}$, $\kappa = 3\ \mathrm{W/(m\cdot K)}$, $E = 250\ \mathrm{GPa}$, $\sigma_0=200\ \mathrm{MPa}$, $\nu=0.23$, $\alpha_\mathrm{th} = 2.2\ \mathrm{ppm/K}$, $\epsilon_\mathrm{rad} = 0.05$, $\alpha_\mathrm{abs}=0.5\ \%$, $h=50\ nm$.}
    \label{fig:SF_mbrn}
    \end{center}
\end{figure}
For the specific case where the temperature is constant within the source region (red dashed curve in Fig.~\ref{fig:SF_mbrn}a), the temperature profile is given by
\begin{equation}\label{eq_T_ecc}
    \Delta T(r) = 
    \begin{cases}
    \frac{P_0}{4 \pi \kappa h} ln\left(\frac{D_\mathrm{eff}^2}{4 w_0^2}\right) & \mathrm{for}\ 0 \leq r < w_0 \\
    \frac{P_0}{4 \pi \kappa h} ln\left(\frac{D_\mathrm{eff}^2}{4 r^2}\right) & \mathrm{for}\ w_0 \leq r \leq \frac{D_\mathrm{eff}}{2}.
    \end{cases}
\end{equation}
For this scenario, an analytical solution is available for the shape factor \cite{Bergman2017}
\begin{equation} \label{eq_Sf_ecc}
s_f(r,\theta,D_\mathrm{eff},w_0)=\frac{2 \pi h}{cosh^{-1} \left (\frac{D_\mathrm{eff}^2+4 w_0^2-4r^2}{4 D_\mathrm{eff} w_0} \right )}.
\end{equation}
The resulting dissipated heat is given by $q = s_f \kappa \Delta T(w_0)$.
For a laser beam impinging on the drumhead with input power $P_0$, the resulting temperature profile is given by (light blue solid curve in Fig~\ref{fig:SF_mbrn}a)
\begin{equation} \label{eq_T_laser}
    \Delta T(r) = 
    \begin{cases}
    \frac{P_0}{4 \pi \kappa h} \left [ \left (1 - \frac{r^2}{w_0^2}\right) + ln\left(\frac{D_\mathrm{eff}^2}{4 w_0^2}\right) \right] & \mathrm{for}\ 0 \leq r < w_0
    \\
    \frac{P_0}{4 \pi \kappa h} ln\left(\frac{D_\mathrm{eff}^2}{4 r^2}\right) & \mathrm{for}\ w_0 \leq r \leq \frac{D_\mathrm{eff}}{2}.
    \end{cases}
\end{equation}
Eq.~\eqref{eq_T_laser} differs from \eqref{eq_T_ecc} within the heated region, due to the different boundary conditions. In this case, 
the corresponding shape factor $s_f$ is obtained by rewriting $\Delta T(w_0)$ as a function of the maximum temperature rise $\Delta T_\mathrm{max}$. For the simple case of concentric, conduction limited heat transport problem, this relation is given by \cite{Kurek2017}
\begin{equation} \label{eq_T}
    \Delta T(w_0)=\frac{P_0}{4 \pi \kappa h} ln \left (\frac{D_\mathrm{eff}^2}{4 w_0^2} \right) = \Delta T_\mathrm{max} - \frac{P_0}{4 \pi \kappa h}.
\end{equation}
Substituing Eq.~\eqref{eq_T} into Fourier's law gives
\begin{equation} \label{eq_P0_Tavg}
    P_0=s_f(r,\theta, D_\mathrm{eff}, w_0) \kappa \Delta T(w_0)=\frac{2 \pi h}{cosh^{-1}\left(\frac{D_\mathrm{eff}^2+4 w_0^2-4 \mathbf{r}^2}{4 D_\mathrm{eff} w_0}\right)} \kappa \left ( \Delta T_\mathrm{max} - \frac{P_0}{4 \pi \kappa h} \right ).
\end{equation}
Rearranging Eq.~\eqref{eq_P0_Tavg} as a function of the peak temperature rise $\Delta T_\mathrm{max}$ gives
\begin{equation} \label{eq_P0_Tmax}
    P_0=\frac{4 \pi h}{2 cosh^{-1}\left(\frac{D_\mathrm{eff}^2+4w_0^2-4 \mathbf{r}^2}{4 D_\mathrm{eff} w_0}\right)+1}\ \kappa\ \Delta T_\mathrm{max} = s_f(r,\theta, D_\mathrm{eff}, w_0) \kappa \Delta T_\mathrm{max}
\end{equation}
Eq.~\eqref{eq_P0_Tmax} describes the heat conduction losses with respect to the maximum temperature rise. The analytical solution \eqref{eq_P0_Tmax} has been tested for different heat source positions against FEM simulations, showing excellent agreement. Fig.~\ref{fig:SF_mbrn}b shows the resulting FEM (solid curves) and analytical (dashed curves) temperature profiles, for an impinging laser power of $10\ \mathrm{\mu W}$ and a beam waist of $1\ \mathrm{\mu m}$. For the implementation of the MTF, the ratio between mean and peak temperature $\beta$ must be found. 
Combining the two expressions of Eq.~\eqref{eq_T_laser}, it is possible to extract the peak temperature
\begin{equation} \label{eq_Tmax}
    \Delta T_\mathrm{max}=\Delta T(0)=\frac{P_0}{4 \pi \kappa h} \left [1+ln \left (\frac{D_\mathrm{eff}^2}{4 w_0^2} \right ) \right ]
\end{equation}
Integrating Eq.~\eqref{eq_T_laser} over the whole resonator area gives the mean temperature
\begin{align} \label{eq_Tavg}
    \braket{\Delta T} = \frac{1}{A} \iint_A \Delta T(r,\theta) dA = \frac{4}{\pi D_\mathrm{eff}^2} \left [ \int_0^{2\pi}\int_0^{w0} \Delta T(r) r dr d\theta + \int_0^{2\pi}\int_{w_0}^{\frac{D_\mathrm{eff}}{2}} \Delta T(r) r dr d\theta \right ] \nonumber \\
    =\frac{4}{\pi D_\mathrm{eff}^2} \frac{P_\mathrm{abs}}{4 \pi \kappa h} \left [\int_0^{2\pi}\int_0^{w0} \left[ -\frac{r^2}{w_0^2} + 1 + ln\left(\frac{D_\mathrm{eff}^2}{4w_0^2}\right) \right]r dr d\theta - \int_0^{2\pi}\int_{w_0}^{\frac{D_\mathrm{eff}}{2}} ln\left(\frac{4r^2}{D_\mathrm{eff}^2}\right) r dr d\theta \right ] 
    \nonumber \\ = \frac{4}{\pi D_\mathrm{eff}^2} \frac{P_\mathrm{abs}}{4 \pi \kappa h} \pi \left[\frac{w_0^2}{2} + w_0^2 ln\left(\frac{D_\mathrm{eff}^2}{4w_0^2}\right) + \frac{D_\mathrm{eff}^2}{4} - w_0^2 - w_0^2 ln\left(\frac{D_\mathrm{eff}^2}{4w_0^2}\right)\right] = \frac{P_0}{4 \pi \kappa h} \left(1 - \frac{1}{2}\frac{4 w_0^2}{D_\mathrm{eff}^2} \right). 
\end{align}
Hence, the $\beta$ factor for the drumhead design can be written as
\begin{equation} \label{eq_beta_mbrn}
    \beta(r,\theta,D_\mathrm{eff},w_0) = \frac{1 - \frac{1}{2}\frac{4 w_0^2}{D_\mathrm{eff}^2}}{1 + ln\left(\frac{D_\mathrm{eff}^2}{4 w_0^2}\right)}\left(1 - \frac{4 r^2}{D_\mathrm{eff}^2}\right),
\end{equation}
with the first term denoting the ratio between mean and maximum temperature rise, while the second term expressing the spatial dependence of the $\beta$ factor. The latter follows by an heuristic approach, by fitting the FEM results.

\subsubsection{Heat localization}

\begin{figure}[h]
    \begin{center}
    \centering
    \includegraphics[width = 1\textwidth]{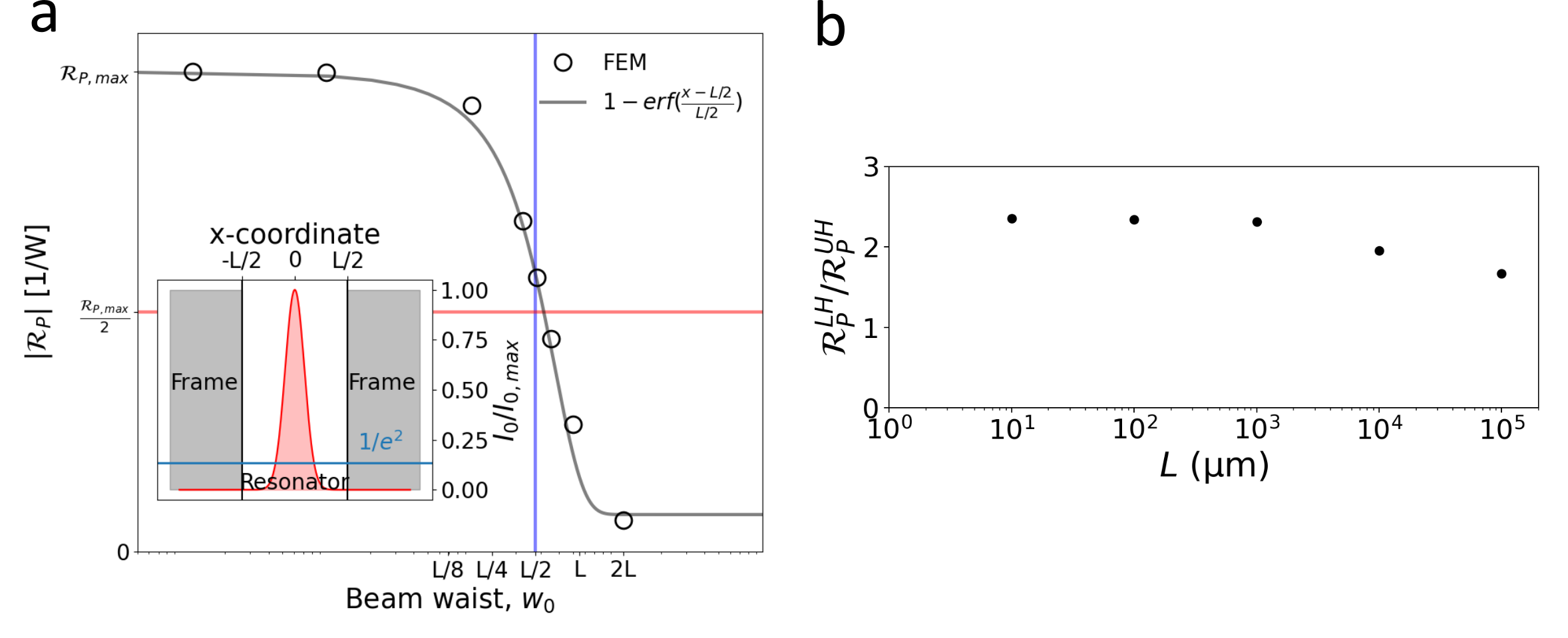}
    \caption{\textbf{Heat localization in membranes.} \textbf{a} FEM-aided relative power responsivity of a membrane resonator of side length $L$, as a function of the light source beam waist $w_0$. A laser with a gaussian profile is defined as heating source, with a constant input power $P_0$ and concentric to the resonator. Inset: gaussian beam profile of the input light source. The $1/e^2$ definition has been used for the beam waist $w_0$. \textbf{b} FEM simulated ratio between LH and UH power responsivity as a function of the drumhead side length $L$. FEM parameters: $\rho = 3000\ \mathrm{kg/m^3}$, $c_p=700\ \mathrm{J/(kg\ K)}$, $\kappa = 3\ \mathrm{W/(m\cdot K)}$, $E = 250\ \mathrm{GPa}$, $\sigma_0=200\ \mathrm{MPa}$, $\nu=0.23$, $\alpha_\mathrm{th} = 2.2\ \mathrm{ppm/K}$, $\epsilon_\mathrm{rad} = 0.05$, $\alpha_\mathrm{abs}=0.5\ \%$, $h=50\ nm$.}
    \label{fig:beta}
    \end{center}
\end{figure}

\begin{figure}[h]
    \begin{center}
    \centering
    \includegraphics[width = 1\textwidth]{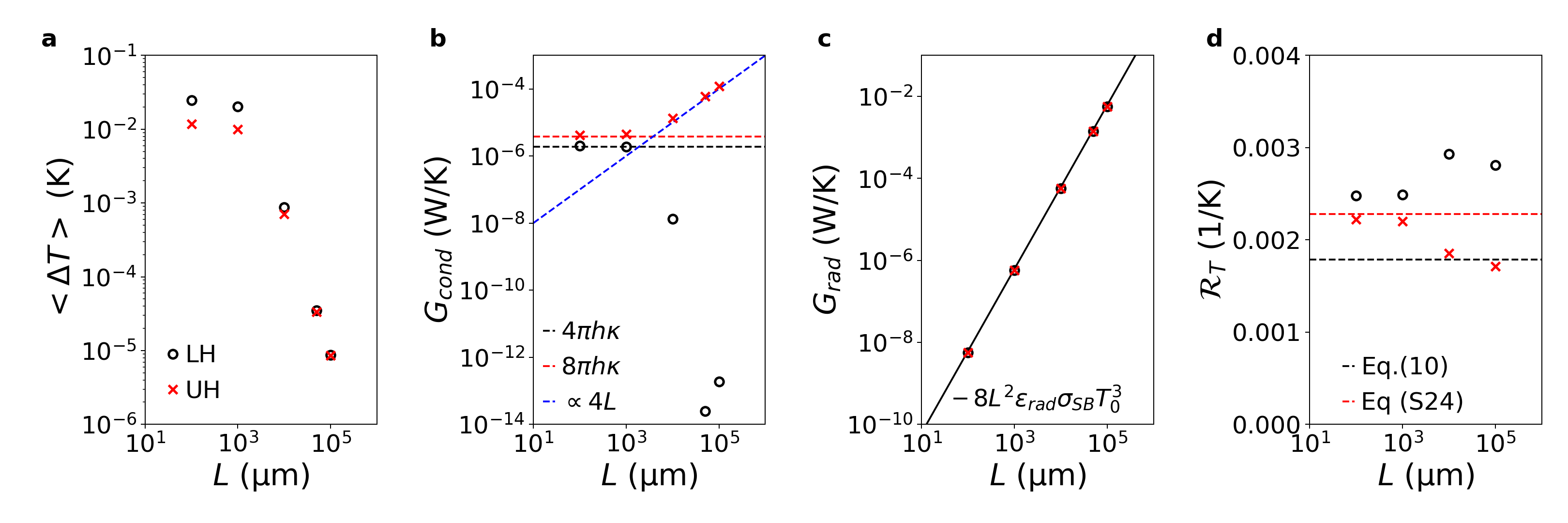}
    \caption{\textbf{Heat localization in membranes.} \textbf{a} FEM simulated mean temperature rise of square drumheads for localized (LH, black circles) and uniform heating (UH, red crosses) \textbf{b} Corresponding thermal conductance due to conduction. The blue curve shows the linear increase in $G_\mathrm{cond}$ as a function of the drumhead perimeter $4 L$ for UH. \textbf{c} Corresponding thermal conductance due to radiation. \textbf{d} Corresponding relative temperature responsivity. FEM parameters: $\rho = 3000\ \mathrm{kg/m^3}$, $c_p=700\ \mathrm{J/(kg\ K)}$, $\kappa = 3\ \mathrm{W/(m\cdot K)}$, $E = 250\ \mathrm{GPa}$, $\sigma_0=200\ \mathrm{MPa}$, $\nu=0.23$, $\alpha_\mathrm{th} = 2.2\ \mathrm{ppm/K}$, $\epsilon_\mathrm{rad} = 0.05$, $\alpha_\mathrm{abs}=0.5\ \%$, $h=50\ nm$.}
    \label{fig:G_Rt_mrb}
    \end{center}
\end{figure}

FEM simulations have also been conducted for the drumhead resonator to show the dependence of the power responsivity on the localization of the heating source. Fig.~\ref{fig:beta}a shows the FEM analysis for a drumhead of side length $L=1$ mm. As the beam waist $w_0$ of a concentric Gaussian beam is increased, $\mathcal{R}_{\mathrm{P}}$ reduces following the complementary error function $1 - erf[(w_0-L/2) / L/2]$, being the absorbed power the convolution between resonator and laser spot size. For tightly focused beams, all the power is concentrated onto the resonator, resulting in a higher mean temperature increase. For a beam waist $w_0 \simeq L/2$, $\mathcal{R}_{\mathrm{P}} \simeq \mathcal{R}_{\mathrm{P,max}} / 2$. This analysis has been carried out also for different drumhead side lengths $L$, as done for the strings. Fig.~\ref{fig:beta}b shows the FEM results. For $L\leq1$ mm, the power responsivity for a localized heating source is $\approx 2\times$ higher than for the uniform heating condition. Again, this means in a two-fold improvement in power responsivity. For larger drumheads, the ratio $\mathcal{R}_P^{LH}/\mathcal{R}_P^{UH}$ reduces for the same reason as in strings.
Substituting Eq.~\eqref{eq_P0_Tmax} and \eqref{eq_beta_mbrn} into equation \eqref{eq_Gcond_mtf} for a uniform ($w_0=D/2$) heating concentric ($r=0$) to the drumhead gives
\begin{equation}
    G_\mathrm{cond} = \frac{4 \pi h \kappa}{1 + 2 cosh^{-1}(1)} \frac{1 + ln(1)}{1 - \frac{1}{2}} = 8 \pi h \kappa,
\end{equation}
meaning that the conductive contribution is doubled. This is clearly show in Fig.~\ref{fig:G_Rt_mrb}b. It is also shown here that $G_\mathrm{cond}$ increases linearly with $L$ for increasingly larger drumheads for uniform heating (dashed blue curve). Indeed, uniformly heated large drumheads will dissipate more heat through the frame. Fig.~\ref{fig:G_Rt_mrb}c shows the corresponding heat radiation. Both uniformly and locally heated drumheads follow the same trend. Conversely, the relative temperature responsivity changes as a function of the heat localization in drumheads. Indeed, the thermal stress is in general function of the spatial temperature distribution. For a circular drumhead, the position-dependent thermal stress can be written as \cite{Kurek2017, Zhang2020}
\begin{equation}
    \sigma(T) = \sigma_0 + \sigma_\mathrm{th}(T) = \sigma_0 \left[1 + \frac{\sigma_\mathrm{th}(T)}{\sigma_0}\right]=\sigma_0 \left\{1 -\alpha_\mathrm{th} \frac{E}{\sigma_0} \left[\frac{1+\nu}{1-\nu}\frac{\braket{\Delta T}}{2} + \frac{1}{r^2}\int_{0}^{r}r'\Delta T(r') dr'\right]\right\}.
\end{equation}
Hence, this thermal stress depends on the temperature profile on the drumhead, as well as the temperature responsivity. For a uniformly distributed temperature, the integral becomes independent of $r$ and equale to $\braket{\Delta T} /2$, leading to $\mathcal{R}_T$ of the form given in Eq.~(10) in the main text. This is the case of large drumheads ($L> 1$ mm) under uniform heating (red crosses), as shown in Fig.~\ref{fig:G_Rt_mrb}d. For $L<1$ mm, the temperature profile cannot be assumed constant anymore, and the temperature responsivity is given by \cite{Kurek2017}
\begin{equation}
    \mathcal{R}_T = -\frac{\alpha_\mathrm{th}}{2(1 - \nu)}\frac{E}{\sigma_0} [2-\nu - 0.642(1-\nu)]
\end{equation}
It can clearly be seen that, the localization (LH, $L>1$ mm) improves the temperature responsivity, leading to a two-fold improvement in the overall power responsivity.

\subsection{Trampoline}
\begin{figure}
    \begin{center}
    \centering
    \includegraphics[width = 0.5\textwidth]{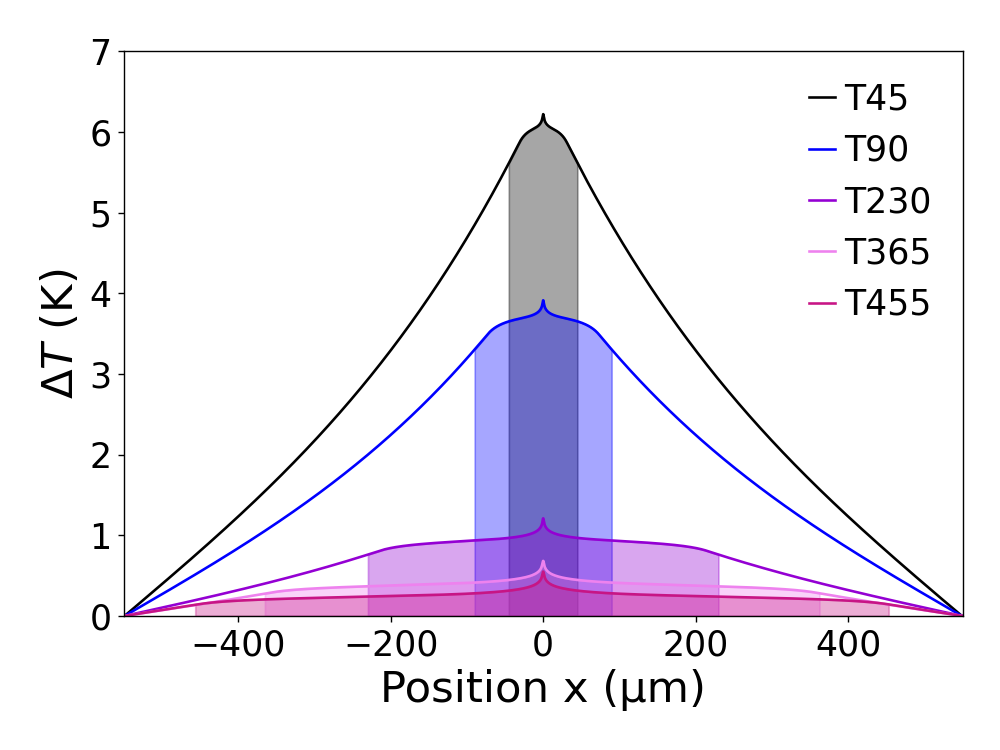}
    \caption{\textbf{Trampoline temperature profile.} FEM simulated temperature distribution along a X-cut line, for the five different trampoline dimensions analyzed experimentally. The shaded regions denote the central areas. Input optical parameters: input power $P_0 = 10\ \mathrm{\mu W}$, beam waist $w_0 = 1\ \mathrm{\mu m}$. FEM parameters: $\rho = 3000\ \mathrm{kg/m^3}$, $c_p=700\ \mathrm{J/(kg\ K)}$, $\kappa = 3\ \mathrm{W/(m\cdot K)}$, $E = 250\ \mathrm{GPa}$, $\sigma_0=200\ \mathrm{MPa}$, $\nu=0.23$, $\alpha_\mathrm{th} = 2.2\ \mathrm{ppm/K}$, $\epsilon_\mathrm{rad} = 0.05$, $\alpha_\mathrm{abs}=0.5\ \%$, $w = 5\ \mathrm{\mu m}$, $h=50\ nm$.}
    \label{fig:Tprofile_trmp}
    \end{center}
\end{figure}

For the trampoline design, only heeating sources impinging onto the central pad are considered ($r \leq L/2$). Heat conduction losses are here constrained along the four tethers, each of length $L_t$, connecting the pad to the frame (see main text). The Fourier law at thermal equilibrium gives
\begin{equation}\label{eq_Fourier_trmp}
    P_0 = \frac{4 w h}{L_t} \kappa \Delta T_\mathrm{max}
\end{equation}
For an illumination only on the central pad, the $\beta$ factor is constant and $\beta = 1$, as supported by FEM simulations. These have been performed for tightly focused Gaussian beam impinging in the center of the five trampolines characterized experimentally in the main text. Fig.~\ref{fig:Tprofile_trmp} displays the corresponding temperature field along a X-cut line. It is composed of a membrane-like temperature distribution within the central pad (shadowed regions), and a linear string-like profile along the tethers. Since the temperature gradient in the central pad is smaller than the gradient at the tethers, and remains constant for different heating source positions, $\beta=1$ (see also Fig.~\ref{fig:ratio_loc_trmp}).

\subsubsection{Heat localization}
\begin{figure}[h]
    \begin{center}
    \centering
    \includegraphics[width = 0.5\textwidth]{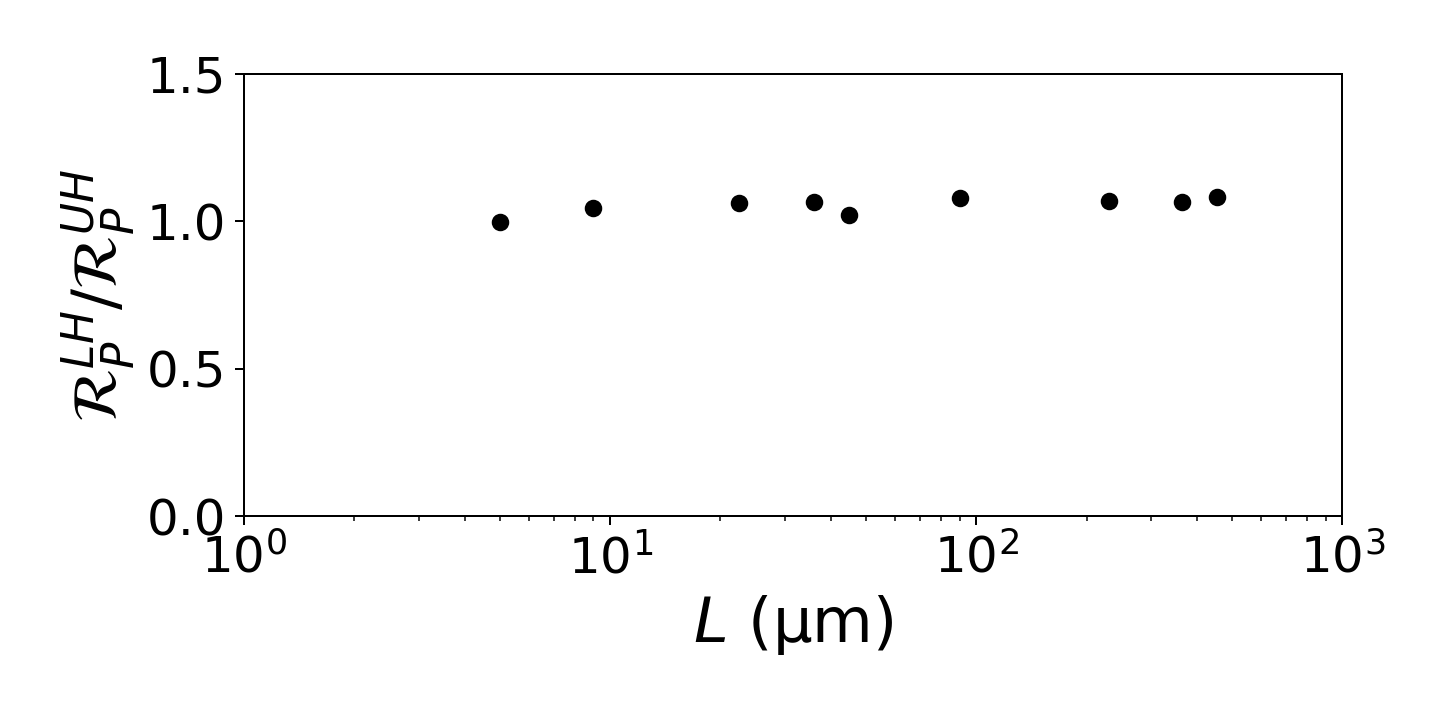}
    \caption{\textbf{Heat localization in trampolines.} Ratio between power responsivity for a localized (LH) and uniform (UH) heating condition. FEM parameters: $\rho = 3000\ \mathrm{kg/m^3}$, $c_p=700\ \mathrm{J/(kg\ K)}$, $\kappa = 3\ \mathrm{W/(m\cdot K)}$, $E = 250\ \mathrm{GPa}$, $\sigma_0=200\ \mathrm{MPa}$, $\nu=0.23$, $\alpha_\mathrm{th} = 2.2\ \mathrm{ppm/K}$, $\epsilon_\mathrm{rad} = 0.05$, $\alpha_\mathrm{abs}=0.5\ \%$, $w = 5\ \mathrm{\mu m}$, $h=50\ \mathrm{nm}$.}
    \label{fig:ratio_loc_trmp}
    \end{center}
\end{figure}

Heating localization has been also studied for trampolines. The greater (lesser) localization of the heat source does not improve (worsen) the overall relative power responsivity, as clearly shown in Fig.~\ref{fig:ratio_loc_trmp}. Here, the ratio between the FEM simulated power responsivity in localized and uniform heating conditions is displayed, showing a value of unity for all the central pad side lengths analyzed here.

Table \ref{tab:shape_factor} shows a summary of the shape $s_f$ and $\beta$ factors for the different designs.
\begin{table*}[h!]
\caption{\label{tab:shape_factor} Shape factor $s_f$ and $\beta$ factor for a square drumhead resonator with side length $L$.}
\begin{ruledtabular}
\begin{tabular}{ccc}
\large Design &\large $s_f(\mathbf{r}, L, w_0)$ &\large $\beta (\mathbf{r}, L, w_0)$\\
\hline
\hline
\large String &\large $\frac{4 h w_{str}}{L-4\frac{\mathbf{r}^2}{L}}$ &\large $\frac{1}{2}$ \\
\hline
\large Drumhead &\large $\frac{4 \pi h}{2 cosh^{-1} \left( \frac{L^2/\pi + w_0^2-\mathbf{r}^2}{2 L w_0/\sqrt{\pi}} \right)  +1}$ &\large $\frac{1-\frac{1}{2} \left( \frac{w_0^2 \pi}{L^2}\right)}{1- ln \left( \frac{w_0^2 \pi}{L^2} \right) } \left(1-\frac{\mathbf{r}^2 \pi}{L^2} \right)$\\ 
\hline
\large Trampoline &\large $\frac{4 w h}{L_t}$ \small for $|r|<L/2$ &\large 1 
\end{tabular}
\end{ruledtabular}
\end{table*}

\section{Experimental Q factors}
\begin{figure}[h!]
    \begin{center}
    \centering
    \includegraphics[width = 0.9\textwidth]{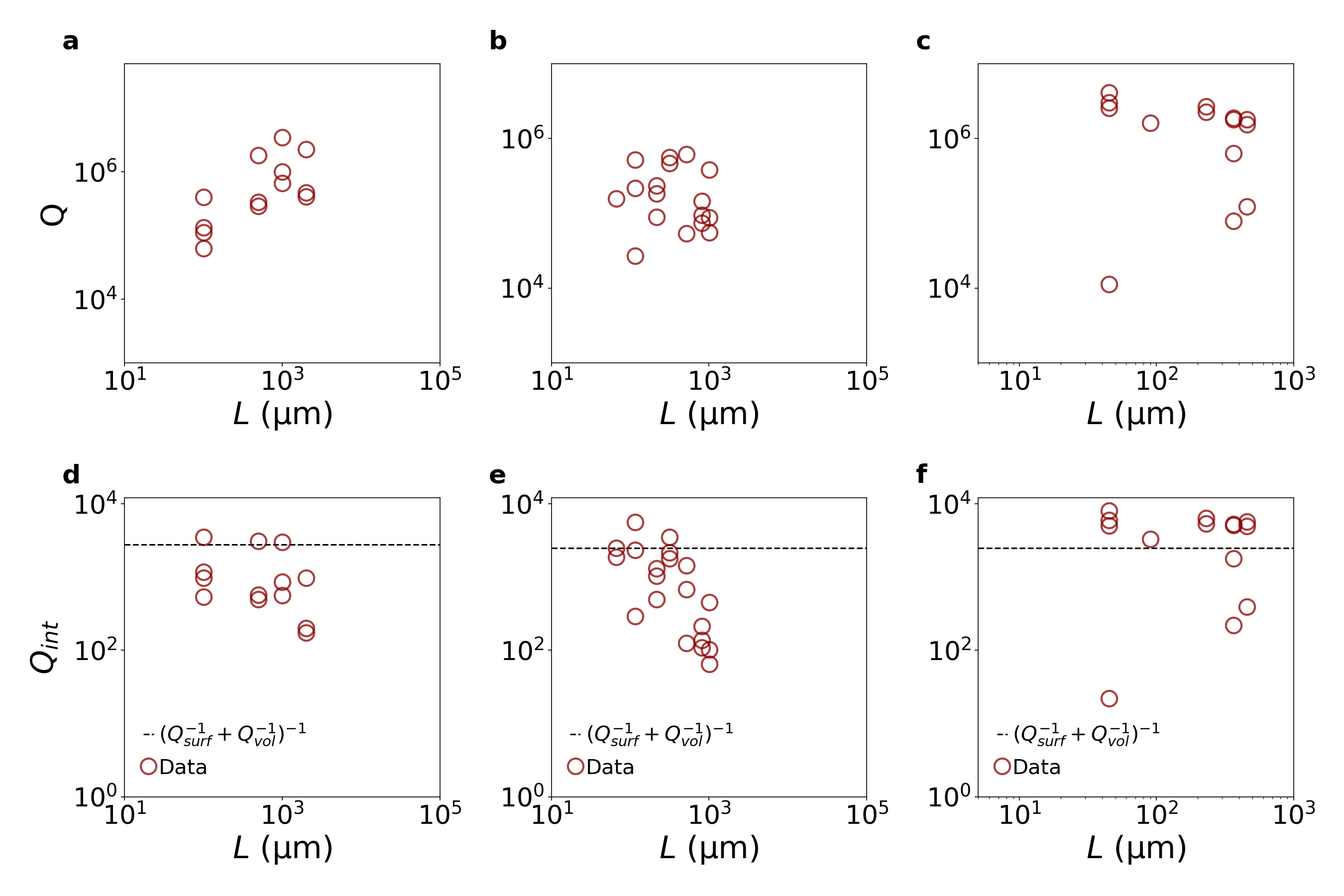}
    \caption{\textbf{Q factors.} Measured Q factor for: \textbf{a} strings, \textbf{b} drumheads, \textbf{c} trampolines. \textbf{d}-\textbf{f} Corresponding intrinsic Q.}
    \label{fig:Q_meas}
    \end{center}
\end{figure}

Their quality factor of each resonator analyzed in the main text has been measured for the theoretical calculations of the AD, due to the $Q$-dependence of the additive phase noise (Eq.~27 in the main text). Fig.~\ref{fig:Q_meas}b displays the experimental values for the damping diluted $Q$ and intrinsic $Q_\mathrm{int}$. The latter results to be mainly dominated by surface losses $Q_\mathrm{surf}$, as expected for this thickness, for all the analyzed structures \cite{Villanueva2014}. Moreover, chip mounting constitute her another source of mechanical dissipation, as observed in the data scattering \cite{Schmid2011}.

\section{Photothermal back-action frequency noise}
To understand the magnitude of the photothermal back-action on the final fractional frequency fluctuations of the resonator, the intensity power spectral density $S_I(\omega, \lambda)$ has been measured recording the optical power in time $P_0(t)$ for $2$ minutes, as explained in \textit{Material and methods}. This optical power is then converted in frequency,
\begin{equation} \label{eq_Rpfit}
    f_0(P_0, t) =\ [1 + \alpha(\lambda)\ \mathcal{R}_\mathrm{P}(\omega)\  P_0(t)]\ f_0(0).
\end{equation}
$f_0(0)$ denotes the resonator eigenfrequency for no impiging optical power ($P_0 = 0\ \mathrm{\mu W}$). Finally, the corresponding AD is computed (dark violet curves in the main text).

\bibliography{nep_biblio.bib}